\documentclass[letterpaper]{article} 
\usepackage{aaai24}  
\usepackage{times}  
\usepackage{helvet}  
\usepackage{courier}  
\usepackage[hyphens]{url}  
\usepackage{graphicx} 
\usepackage{natbib}  
\usepackage{caption} 
\usepackage{algorithm}
\usepackage{algorithmic}

\usepackage{newfloat}
\usepackage{listings}
\DeclareCaptionStyle{ruled}{labelfont=normalfont,labelsep=colon,strut=off} 
\floatstyle{ruled}
\newfloat{listing}{tb}{lst}{}
\floatname{listing}{Listing}
\usepackage{amssymb,amsmath,amsthm}
\usepackage{multirow}
\usepackage{siunitx}
\usepackage{tabularx}
\usepackage{wasysym}

\newcommand{\norm}[1]{\,\left\Vert {#1} \right\Vert\,}

\newcommand\Tstrut{\rule{0pt}{2.2ex}}         
\newcommand\Bstrut{\rule[-1.3ex]{0pt}{0pt}}   

\newtheorem{optim}{Optimization-problem}

\newcommand{\covidscore}{\textsc{covidscore}}
\newcommand{\openabm}{OpenABM}
\newcommand{\covasim}{Covasim}
\newcommand{\covidn}{COVID19}

\usepackage{xcolor}         
\definecolor{green}{RGB}{66, 180, 20}
\definecolor{red}{RGB}{219,68,55}
\definecolor{blue}{RGB}{66,133,244}
\definecolor{magenta}{RGB}{200, 90, 200}

\title{Protect Your Score: Contact-Tracing \\  with Differential Privacy Guarantees}
\author {
    Rob Romijnders\textsuperscript{\rm 1},
    Christos Louizos\textsuperscript{\rm 2},
    Yuki M. Asano\textsuperscript{\rm 1},
    Max Welling\textsuperscript{\rm 1}
}
\affiliations {
    \textsuperscript{\rm 1}University of Amsterdam \\
    \textsuperscript{\rm 2}Qualcomm AI research\\
    r.romijnders@uva.nl
}

\usepackage{bibentry}

\begin{document}

\maketitle

\begin{abstract}
The pandemic in 2020 and 2021 had enormous economic and societal consequences, and studies show that contact tracing algorithms can be key in the early containment of the virus. While large strides have been made towards more effective contact tracing algorithms, we argue that privacy concerns currently hold deployment back. The essence of a contact tracing algorithm constitutes the communication of a risk score. Yet, it is precisely the communication and release of this score to a user that an adversary can leverage to gauge the private health status of an individual. We pinpoint a realistic attack scenario and propose a contact tracing algorithm with differential privacy guarantees against this attack. The algorithm is tested on the two most widely used agent-based \covidn\ simulators and demonstrates superior performance in a wide range of settings. Especially for realistic test scenarios and while releasing each risk score with $\varepsilon=1$ differential privacy, we achieve a two to ten-fold reduction in the infection rate of the virus. To the best of our knowledge, this presents the first contact tracing algorithm with differential privacy guarantees when revealing risk scores for \covidn. 
\end{abstract}

\section{Introduction} \label{sec:introduction}
The COVID19 pandemic had enormous economic and societal consequences~\cite{societal_impact_01,societal_impact_02,econ_impact_01,berger2020seir}. Some sources estimate the global economic impact at more than a trillion US dollars~\cite{econ_impact_02}. Previous studies show that contact tracing apps can aid understanding and mitigate the early rise of the pandemic~\cite{covi,baker2021epidemic,crisp,perra2021non}. Most studies, however, focused on the effectiveness of the pandemic mitigation, while we argue that privacy concerns hold the deployment back~\cite{raskar2020apps,covi}. Population surveys during and after the pandemic show that mistrust and `worries about privacy' are among the top three reasons not to use a contact tracing app~\cite{privacy_concern_01,privacy_concern_02,privacy_concerns_03}.

Several studies argue for the studying of privacy in contact tracing algorithms~\cite{tracing_controversy_sk,tracing_controversy_mobile,dyda2021differential}.  We also quote an influential journal stating that most individuals would consider the privacy risks \textit{``to be unacceptably high''}~(The Lancet; \citeauthor{bengio2020need}~\citeyear{bengio2020need}). Yet, to the best of our knowledge, no research has been published on differential privacy when releasing a risk score to a user for the purpose of contact tracing. \looseness=-1

Despite security measures, a contact tracing algorithm needs to assign a risk score and release the score either directly to the user or indirectly by the signal to get tested. It is precisely the communication and release of this score that an adversary can leverage to gauge the private health status of an individual. For the rest of the paper, we refer, by the name \covidscore, to a risk score that a contact tracing algorithm assigns to a user and communicates to other users. Most papers about privacy and security aspects during the \covidn\ pandemic center on security measures~\cite{security01} for establishing contacts, where approaches such as hashing were studied for establishing contacts~\cite{security02,security03}. In this work, we assume that these security precautions are adhered to and, in the presence of the security measures, identify another privacy attack on the \covidscore:

\begin{quote}
An adversary wants to determine the \covidscore\ of a victim. The adversary installs the app and only makes contact with the victim. The next day, the adversary observes a change in their  \covidscore. This change is due to the victim, and the adversary reconstructs the \covidscore\ of the victim.
\end{quote}

The naïve approach of simply adding noise to any revealed \covidscore\ does lead to increased uncertainty at the adversary about the score of a victim. 
However, adding noise naturally decreases the utility of a contact tracing algorithm. 
To address this conundrum, we propose a novel algorithm that, while adding noise, maintains good results in mitigating a peak of the pandemic. 
Moreover, we prove \textit{differential privacy}~\cite{dwork_dp} for the release of the \covidscore\ and demonstrate strong performance even when $\varepsilon\leq1$ per message.

In this paper, we make the following contributions:

\begin{enumerate}
  \item We concretize a privacy attack in contact tracing with important implications, and we propose a novel decentralized algorithm with a differential privacy guarantee against this attack. To the best of our knowledge, we are the first to study the differential privacy of a \covidscore\ on top of the standard security measures.
  \item The trade-off between privacy and utility is studied on two widely used simulators. The method is compared against existing methods for differential privacy, and show favourable results on the privacy-utility trade-off. For the case $\varepsilon=1$, we show that up to the million scale, our algorithm achieves a two to ten times smaller infection rate, compared to traditional contact tracing.
  \item To evaluate our algorithms' robustness across a range of realistic conditions, we also evaluate our algorithm in two challenging circumstances: under imperfect tests for \covidn\ and a reduced test protocol.
\end{enumerate}

The code for our method and all experiments is available at \texttt{github.com/RobRomijnders/dpfn\_aaai}.

\section{Related Work}\label{sec:related_work}

This section discusses the related work for our method. We discuss the current agent-based statistical contact tracing approaches and the recent research in privacy for COVID containment strategies.

\textit{Statistical contact tracing: } Various approaches have been published about statistical contact tracing, especially during the \covidn\ pandemic. 
\citet{burdinski2022understanding} test the efficacy of traditional contact tracing and run simulations, including self-isolation strategies. \citet{li2021impact} use a message-passing approach and analyze an isolation policy based on risk-score estimation. \citet{crisp} investigate statistical contact tracing using Gibbs sampling and show results on a simulator based on stochastic block models. \citet{braunstein2023small} propose an inference model similar to belief propagation but do not test on \covidn\ simulators.
Most similar to ours, \citet{baker2021epidemic} propose statistical contact tracing using belief propagation on a collapsed graph. \citet{romijnders2022notimetowaste} propose another algorithm for statistical contact tracing, improving over the previous approach and comparing statistical contact tracing under constrained communication.

\textit{Privacy in \covidn\  containment strategies: } During the pandemic, many papers raised concerns about privacy and security in contact tracing. The first step is the design of decentralized algorithms where no central entity has the \covidscore\ of multiple individuals~\cite{baker2021epidemic,crisp,romijnders2022notimetowaste}. Yet many security issues remain. \citet{central_controversy_app} provides an overview of the methods for proximity tracing and its various threat models. A paper in Nature Communications highlights the pitfalls of collecting such data from smartphones~\cite{tracing_controversy_mobile} and calls for more research in privacy. Obtaining the contact graph and the various threat models for sharing GPS location are discussed in papers such as~\cite{raskar2020apps,security01,security02,security03}. Examples of approaches that study obtaining contacts under secure and private circumstances are~\cite{decent_example_singapore,decent_example_pact,DBLP:journals/corr/abs-2003-11511}.

\textit{Differential privacy in decentralized inference: } For the general purpose of statistical inference, a few but existent papers have studied differential privacy (DP).
For example, DP for MCMC has been studied~\cite{yildirim2019exact,heikkila2019differentially}. We implement and compare to a method~\cite{privacyforfree,DBLP:conf/uai/FouldsGWC16} that specifically tailors to Gibbs sampling~\cite{crisp}. \citet{DBLP:journals/tods/ZhangCPSX17} analyzes the computation of marginals in a message-passing approach for inference. However, that paper uses the Laplace mechanism for dealing with real-valued random variables of fixed dimensionality. In contrast, our random variables are discrete-valued and have varying degrees.
Like us, \citet{zou2015belief} use the local structure of the belief propagation message to obtain a privacy guarantee. However, they consider a different form of privacy and do not study differential privacy. \looseness=-1

Two noteworthy approaches in contemporary literature study DP in the context of \covidn, but both methods do not relate to contact tracing. \citet{vadrevu2020hybrid} focuses on collecting and clustering medical records and does not mention contact tracing; \citet{vepakomma2021dams} focuses on collecting user trajectories with DP guarantees. However, these works are vulnerable to the same attack we study in this paper.

For other approaches to privacy concerning the release of a \covidscore, previous research has mentioned low-bit quantization of the decentralized messages~\cite{covi,apple_google_app,romijnders2022notimetowaste}. Still, these approaches have no formal guarantee pertaining to privacy. To the best of our knowledge, we are the first paper to propose an algorithm for statistical contact tracing with differential privacy guarantees.

\section{Method}\label{sec:method}

This method section proceeds as follows: first, we explain the model for statistical contact tracing. We discuss three existing approaches for obtaining differential privacy, which will be compared in the experimental section. Then, we propose a composite scheme for differential privacy using a recent message-passing method.

\subsection{Model}
We first present background on the statistical model. Both methods in the later method section use this formulation for the statistical model. This section largely follows notation from previous works~\cite{crisp, romijnders2022notimetowaste,koller_pgm}. 

Every user on every day is modeled as a random variable that takes on one of four states, ${S,E,I,R}$. These states abbreviate for Susceptible, Exposed, Infected, and Recovered~\cite{kermack1927contribution,anderson1992infectious}. This random variable is written as $z_{u,t}$ for user $u$, at time step $t$. The data set of observations is $D_\mathcal{O} = \{o_{u_i,t_i}  \}_{i=1}^O$, which are $O$ observations, each with an outcome $\{0,1\}$ for user $u_i$ at time step $t_i$. 

Test outcomes may have false positive or false negative results, with False Positive rate $\beta$ (FPR) and False Negative rate $\alpha$ (FNR). The model uses the observation distribution:
\begin{equation}
    P\left(o_{u,t}|z_{u, t}\right) =  
    \begin{cases}
    \alpha & \textrm{if}\ z_{u,t}=I \wedge o=0 \\
    1-\alpha & \textrm{if}\ z_{u,t}=I \wedge o=1 \\
    1-\beta & \textrm{if}\ z_{u,t}\in\left\{ S,E,R\right\} \wedge o=0 \\
    \beta & \textrm{if}\ z_{u,t}\in\left\{ S,E,R\right\} \wedge o=1  
    \end{cases}. \label{eqn:test_outcome}
\end{equation}

The random variables $z_{u,t}$ are connected in two directions: over time, the variables evolve in a Markov chain $S \rightarrow E \rightarrow I \rightarrow R$; between users, a contact can influence the transition probability between states. Both interactions are summarized in the Markovian state transition:

\begin{alignat}{2}\label{eqn:crisp_dynamics}
  P(z_{u,t+1}|z_{u,t}, z_{N(u,t)})  = \begin{cases}
    \psi(u, t, z_{N(u,t)})    &  S \rightarrow S\\
    1-\psi(u, t, z_{N(u,t)})  &  S \rightarrow E\\
    1-g                       &  E \rightarrow E\\
    g                         &  E \rightarrow I\\
    1-h                       &  I \rightarrow I\\
    h                         &  I \rightarrow R\\
    1                         &  R \rightarrow R\\
    0                         & \mbox{otherwise}
    \end{cases} \,
\end{alignat}

Here $\psi(\cdot)$ constitutes a noisy-OR model~\cite{koller_pgm} that depends on states of other users:

\begin{align}\label{eqn:crisp_noisy_or}
    \psi(u, t, z_{N(u,t)}) = (1-p_0) (1-p_1)^{|\{(v,u,t) \in D_c: z_{v,t}=I \}|}.
\end{align}

Here $g$, $h$, $p_0$, and $p_1$ are scalar model parameters, and they are set equal to the values from previous literature~\cite{romijnders2022notimetowaste,crisp}. We highlight all parameter settings in Appendix~\ref{app:experimental_details}. $z_{N(u,t)}$ is the set of random variables of all contacts of user $u$ at time step $t$. The data set of contacts, $D_c$, consists of a set of tuples $\{ (u, v, t) \}$, where user $u$ had a (directed) contact with user $v$ at time step $t$. Equation~\ref{eqn:crisp_noisy_or} can be interpreted as a noisy-OR model, where every infected contact decreases the probability of remaining in $S$ state.

\subsection{DP Contact Tracing Methods }\label{sec:dp_existing}

We introduce three methods to obtain a differentially private \covidscore, as defined in the attack model. We follow the conventional definition of $(\varepsilon,\delta)$ differential privacy~\cite{dwork_dp,DBLP:conf/csfw/Mironov17} that says for every $\varepsilon > 0$, $\delta \in [0, 1]$, a mechanism $f(\cdot)$, for any outcome $\Phi$ in the range of $f(\cdot)$, and any two adjacent data sets $D$, $D'$ that differ in at most one element, satisfy the following constraint:

\begin{align}
    p(f(D) \in \Phi) \leq e^\varepsilon p(f(D') \in \Phi) + \delta \label{eqn:definition_dp}
\end{align}

We define two data sets as adjacent when the \covidscore\ of one contact differs between the data sets. The sensitivity, then, is the largest value change of a function between adjacent data sets:


\begin{equation}
\Delta \geq \max_{\{(D,D') : d(D,D') = 1 \}} \norm{f(D) - f(D')}. \label{eqn:defn_neighbor}
\end{equation}

Distance $d()$ is defined as the Hamming distance $d(D,D') = \sum_i \mathbf{1}[D_i \neq D'_i]$, where $\mathbf{1}[\cdot]$ is the indicator function and $D_i$ is one of the contacts' \covidscore. When the sensitivity of function $f()$ is bounded, a common mechanism is to add Gaussian noise. The Gaussian mechanism of~\cite{dwork_dp,balle2018gauss} prescribes the noise variance for a particular sensitivity value,  $\varepsilon$ and $\delta$. \\

We discuss three baseline approaches for experimental comparison: one based on traditional contact tracing, one based on previously studied Gibbs sampling, and one based on noising individual messages regardless of application.

\textbf{Traditional contact tracing. }In traditional contact tracing, users would test themselves when one of their recent contacts has tested positive. Many countries used this policy in the COVID19 pandemic~\cite{baker2021epidemic}. We implement this as a function that calculates the number of positive-testing contacts. If a positive test corresponds to 1 and a negative test corresponds to 0, this function has a sensitivity of 1, according to the definition in Equation~\ref{eqn:defn_neighbor}. We use the (Analytical) Gaussian mechanism accordingly and release its output as the \covidscore~\citep{dwork_dp,balle2018gauss}.

\textbf{Gibbs sampling.} Previous work proposed Gibbs sampling to estimate the \covidscore\ in decentralized contact tracing~\cite{crisp}. For achieving DP, we use an existing method with $\varepsilon$-DP for a sample from a probability distribution with clipped likelihoods~\cite{privacyforfree,DBLP:conf/uai/FouldsGWC16}. Inference for the model specified in Equation~\ref{eqn:crisp_dynamics} makes estimates with Monte Carlo samples from a Gibbs chain. Therefore, if Gibbs samples were obtained under differential privacy, then the Monte Carlo estimate is DP by post-processing. The method provides an $\varepsilon$-DP, which is stronger than the $(\varepsilon,\delta)$-DP of the other methods. 

The Gibbs sampler has two hyperparameters: the value of $B$ to clip the likelihoods and the number of Gibbs samples to draw. We found a value of $B=10$ to work best. Determining the number of Gibbs samples constitutes a topic by itself~\cite{DBLP:books/sp/RobertC04}. Our case is even more complex as each additional sample improves the statistical estimate, but simultaneously increases the privacy bound. We find that taking 10 samples with 10 skip steps, after 100 burn-in steps, works best~\cite{DBLP:books/sp/RobertC04}; taking more samples would worsen the privacy bound, and taking fewer samples worsens the estimate for the \covidscore.

\textbf{Per-message differential privacy.} As a third baseline, we compare against a form of differential privacy at the single message that is communicated between contacts, regardless of the contact tracing algorithm. As the message is a numerical value in the range $[0, 1)$,  one can noise this message and consider the message-passing algorithm to be DP by the post-processing property~\cite{dwork_dp}. To message-passing algorithms are belief propagation~\cite{crisp} and Factorised Neighbors~(FN, \citet{romijnders2022notimetowaste}). We use the latter method as a previous work shows that FN works better for these SEIR models~\cite{romijnders2022notimetowaste}.

Dealing with the constraint that messages are in $[0,1)$, we add noise in the logit domain. A message in the $[0,1]$ domain corresponds to a message in $\mathbb{R}$, transformed by the logit transform $x = \log \frac{y}{1-y}$; and being calculated from the sigmoid function $y = \frac{1}{1+e^{-x}}$. If the messages are clipped to $[\gamma, 1-\gamma]$, then the sensitivity of the mechanism in the logit-domain is $2 \left|\mathrm{logit}(\gamma)\right|$. We use the (Analytical) Gaussian mechanism with this sensitivity and report results for the corresponding $\varepsilon$ and $\delta$ values.

\subsection{Differentially Private Factorized Neighbors}\label{sec:dp_log_normal}

\begin{figure}
    \centering
    \includegraphics[width=0.99\linewidth]{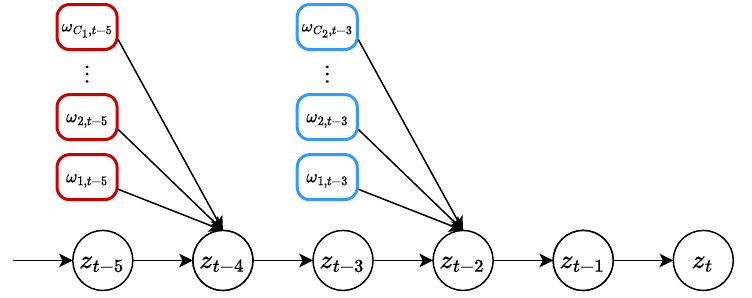}
    \caption{Example of a contact graph. This user has $C_1$ contacts at five time steps in the past and $C_2$ contacts at three time steps in the past. The released \covidscore\ is the estimate of being in state $I$ on time step t. Appendix~\ref{app:fn_traces_explicit} generalizes the method for a general contact graph.}
    \label{fig:contact_graph_3_5}
\end{figure}

In contrast to the previous methods, we now use the structure of Factorized Neighbors (FN, ~\cite{DBLP:conf/uai/Rosen-ZviJY05,romijnders2022notimetowaste}) to propose a novel algorithm. FN is a decentralized approximate inference method that calculates daily a \covidscore, which represents the belief that the user is in the infected state the next day. The update equations for the model in Equation~\ref{eqn:crisp_dynamics} were introduced in~\cite{romijnders2022notimetowaste}. This section analyzes the update equations and proposes a differential privacy method based on composite inputs named differentially private factorized neighbors (DPFN).

We analyze an example for one user and limited contacts here and generalize the method in Appendix~\ref{app:fn_traces_explicit}. Consider revealing the \covidscore, $\phi_{u,t}$ for user $u$ at day $t$. Let's say this user had $C_{1}$ contacts five days before and $C_{2}$ contacts three days before. Figure~\ref{fig:contact_graph_3_5} presents an example of the corresponding contact graph. We rewrite $\omega_{u,t} = 1-p_1 \phi_{u,t}$ for reasons that will become clear shortly. Each contact, $c$, sends a message $\omega_{c,t} \in [0,1]$ to user $u$, and this user calculates their \covidscore. This version of the update equation will be referred to as $F_1(\cdot)$: 

\begin{align}
  \phi_{u,t} = F_1(&\omega_{1,t-5}, \omega_{2,t-5}, \cdots, \omega_{{C_{1}}, t-5}, \nonumber \\ 
    & \quad \omega_{1, t-3}, \omega_{2, t-3}, \cdots, \omega_{{C_{2}}, t-3}) .\label{eqn:fn_before_needs_only_product}
\end{align}

$F_1$ makes a prediction as a function of $C_{1}+C_{2}$ individual messages. However, when analyzing the update equations, the function $F_1$ only depends on messages that appear in a product term.  Then one could rewrite the FN method to:

\begin{align}
    \phi_{u, t} = F_2(\prod_{i=1}^{C_{1}} \omega_{i, t-5}, \quad  \prod_{i=1}^{C_{2}} \omega_{i, t-3} ) \label{eqn:fn_needs_only_product}
\end{align}

FN in the form of $F_2$ only depends on a product of messages. For this reason, we write $\omega_{u,t} = 1-p_1 \phi_{u,t}$. Thus, once such product is modified to have DP, the function $F_2$ will be private by the post-processing property~\cite{dwork_dp}. This was an example for two days, and in Appendix~\ref{app:fn_traces_explicit}, we prove that this decomposition holds for any number of days. \looseness=-1

To derive a bound like Equation~\ref{eqn:definition_dp}, we use a log-normal noise distribution for each message $\omega_{c,t}$, as the family of log-normal distributions is closed under multiplication. The log-normal distribution has a closed-form expression for its R\'enyi divergence, and we will prove DP via R\'enyi differential privacy (RDP,~\cite{DBLP:conf/csfw/Mironov17}). 

A bound on the R\'enyi divergence can be converted to the $\varepsilon$ and $\delta$ for DP~\cite{DBLP:conf/csfw/Mironov17}. As such, we aim to bound the R\'enyi divergence between the two log-normal distributions that correspond to two adjacent data sets, and convert to $(\varepsilon,\delta)$-DP later. For any two log-normal distributions, $p_u$ and $p_v$, with mean parameters $\mu_u$ and $\mu_v$, and with equal variance parameter $\sigma^2_* = C \sigma^2$, the R\'enyi divergence is the following. We assume a product of $C$ messages, each with a variance parameter $\sigma^2$. A detailed derivation is in Appendix~\ref{app:detailed_RDP}, where we also highlight the difference between this method and the Gaussian mechanism.

\begin{align}
    D_{a}(p_u \vert p_v) &=  \frac{a}{2 C  \sigma^2 } (\mu_u - \mu_v)^2 \label{eqn:div_bound_01}
\end{align}

The divergence in Equation~\ref{eqn:div_bound_01} decreases with the number of contacts $C$. So, the more contacts on a day, the smaller the divergence. It remains to upper bound the worst case of $(\mu_u - \mu_v)^2$ for any two adjacent data sets. In Appendix~\ref{app:detailed_RDP} we show that for any two adjacent data sets, $(\mu_u - \mu_v)^2 \leq (\log(1 - \gamma_u p_1) - \log(1 - \gamma_l p_1))^2$. This bound is achieved by clipping every \covidscore\ of the FN computation in the interval $[\gamma_l, \gamma_u]$. Parameter $p_1$ is a model parameter representing the probability that, given a contact, the virus transmits from user to user. Denoting the worst-case divergence in Equation~\ref{eqn:rdp_log-normal} by $\rho$, we have a bounded R\'enyi divergence if the following holds: 
\begin{align}
    \sigma^2 \geq \frac{a}{2C  \rho  }(\log(1 - \gamma_u p_1) - \log(1 - \gamma_l p_1))^2.  \label{eqn:bound_variance_rdp}
\end{align}

Equation~\ref{eqn:bound_variance_rdp} shows that more noise should be added whenever wider clipping values are used or when a user has fewer contacts. Experimentally, we find that tuning the clipping values could slightly improve the results, but another hyperparameter increases the complexity of the method. Therefore, we run all experiments with $\gamma_l=0$ and $\gamma_u=1$.

Algorithm~\ref{alg:dpfn} summarizes the steps in calculating the \covidscore\ with DPFN. The sample is from a log-normal distribution with the parameter $\mu = \omega_{*,t} - \frac{\sigma^2}{2}$. The variance parameter $\sigma^2$ follows from Equation~\ref{eqn:bound_variance_rdp} with $a$, $\rho$, and the specified number of contacts $C = |N(u,t)|$.

\begin{figure}
    \centering
    \includegraphics[width=0.99\linewidth]{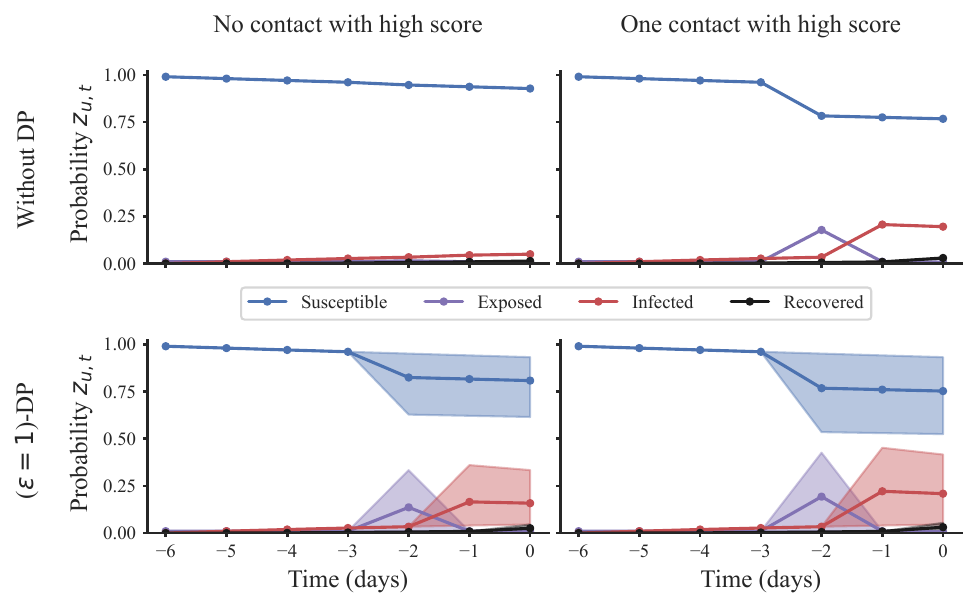}
    \caption{Showing the effect of differential privacy on the approximate inference from FN. An example user has two contacts. Both contacts have a low \covidscore in the left column, while in the right column, one contact has a high \covidscore. The red-shaded region indicates the 20-80 quantiles for sampling the \covidscore\ from the DPFN mechanism of Algorithm~\ref{alg:dpfn}. The regions overlap, which reflects privacy, but the median red line is higher in the right column, which indicates a possible infection and could inform a testing policy.}
    \label{fig:fn_analysis_score}
\end{figure}

\textbf{Visual example.} We illustrate the effect of differential privacy on the estimates for the \covidscore\ by FN. In this example, a user has two contacts at day $-5$. In the left column of Figure~\ref{fig:fn_analysis_score}, both contacts have a low \covidscore; in the right column, one contact has a high \covidscore. The red line indicates the median estimate of being infected, i.e. the \covidscore, and the shaded region indicates the 20-80 quantiles. Interpreting the definition of DP in Equation~\ref{eqn:definition_dp}, changing the score of a single contact should not change the likelihood of an output \textit{too much}. Whether the user has no contact with a high score (left column), or a single contact with a high score (right column), the shaded red regions in the figure overlap, which gives the contact plausible deniability against a potential adversary~\cite{dwork_dp}. The red median line, though, runs slightly higher in the right column, which is the utility needed for a contact tracing algorithm. Naturally, the added noise gives rise to a trade-off, where more noise increases privacy but decreases the utility for a subsequent testing policy and mitigation of a pandemic~\cite{dinur2003revealing}. We address this trade-off in Section~\ref{sec:results}. \looseness=-1

\textbf{Optimize parameters in RDP.} The bound in Equation~\ref{eqn:bound_variance_rdp} uses RDP, and depends on $(a,\rho)$. Yet, we want to report $(\varepsilon, \delta)$-DP.  Previous literature optimizes for the optimal value $a$ via a line search~\cite{abadi2016deep}. Fortunately, for our particular problem, we find a closed-form solution for optimal $a$ and $\rho$, outlined in Appendix~\ref{app:a_opt}. We arrive at the expression for the order $a$ of RDP:
\begin{align} \label{eqn:fixed_point_a}
    a &=  1 + \frac{d + \sqrt{d (d+\varepsilon)}}{\varepsilon} \\ 
    \rho &= \varepsilon - d (a-1)^{-1}
\end{align}

with $d = \log \frac{1}{\delta}$.

\textbf{Assumptions on the algorithm: } The inference runs for a specific time window, $t-T, t-T+1, \cdots, t-1, t$, and an estimate for the probability $p(z_{u,t} = I)$ is released to the user (i.e. using Gibbs sampling or FN). Only this \covidscore\ is released to the user under DP, and inference is run unmodified, in an encrypted space, such as a trusted execution environment\cite{sabt2015trusted}. The differential privacy holds with respect to the message of a contacted user at the time step of the contact. If the user has no other contacts than an adversary, an adversary could gain more information through repeated contacts. In the worst case for $K$ repeated contacts, the differential privacy parameters $\varepsilon$ and $\delta$ increase $K$ fold~\cite{dwork_dp}. We aim to investigate advanced composition bounds for this case in future work~\cite{user_level_privacy}. We assume to have access to a known contact graph, using methods as mentioned in the Related Work in Section~\ref{sec:related_work}. Finally, the attack outlined in the introduction assumes that the adversary uses a contact tracing app and does not want to get infected. Otherwise, a \covidn\ test would reveal the health status. \looseness=-1

\begin{algorithm}[tb]
\caption{Differentially private factorized Neighbors}
\label{alg:dpfn}
\textbf{Input}: Dataset of contacts' \covidscore\, $D = \{ \phi_{c,t} \}$ for all contacts $c,t$ of user $u$ in the set of neighbors $N(u,t)$ \\
\textbf{Parameter}: Privacy parameters $(\varepsilon, \delta)$, model parameters are omitted for clarity\\
\textbf{Output}: \covidscore\ for this user \\ 
\begin{algorithmic}[1] 
\STATE Convert $(\varepsilon,\delta)$-DP parameters to $(a, \rho)$-RDP parameters using Equation~\ref{eqn:fixed_point_a}
\STATE Convert each $\phi_{c,t}$ to $\omega_{c,t}$ using Equation~\ref{eqn:phi_to_S}
\FOR{$t=-T, -T+1, \cdots, -1$}
\STATE $\omega_{*,t} = \prod_{c \in N(u,t)} \omega_{c,t}$
\STATE Calculate $\sigma^2(a, \rho, |N(u,t)|)$ using Equation~\ref{eqn:bound_variance_rdp}
\STATE Calculate $\mu(\omega_{*,t}, \sigma^2)$ using Equation~\ref{eqn:rdp_detailed_sum}
\STATE $\tilde{\omega}_{*,t} \sim \operatorname{log-normal}(\mu, \sigma^2)$
\ENDFOR
\STATE \textbf{return} $F_2( \tilde{\omega}_{*,t=-T}, \tilde{\omega}_{*,t=-T+1}, \cdots, \tilde{\omega}_{*,t=-1})$
\end{algorithmic}
\end{algorithm}

\section{Experiments and Results}\label{sec:results}

We test the differentially private contact tracing algorithms on two widely used simulators. We will explore the trade-off between privacy and utility in two experiments.

\begin{figure*}[th]
    \centering
    \includegraphics[width=.99\linewidth]{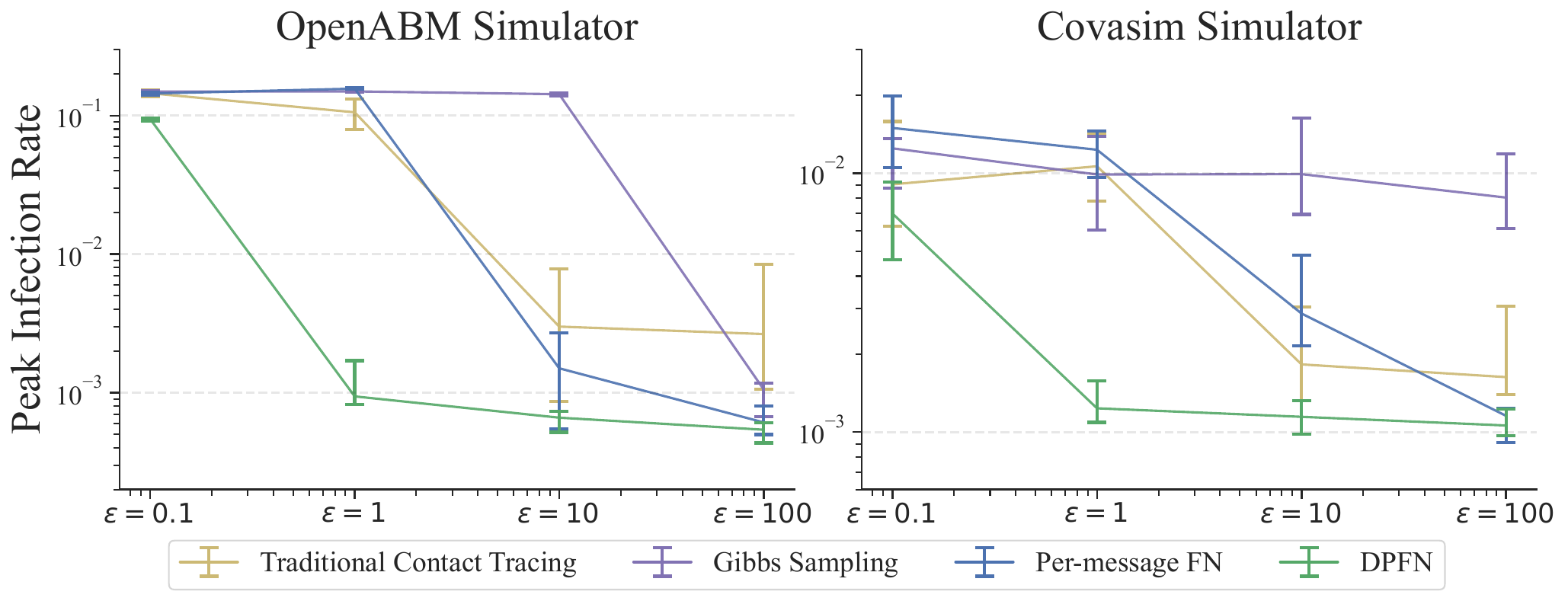}
    \caption{The privacy-utility trade-off for differentially private contact tracing. The y-axis indicates the Peak Infection rate, where lower is better. At $\varepsilon=1$, a common setting for differential privacy, DPFN achieves a lower peak infection rate than all other methods. \openabm\ and \covasim\ are the two most widely used simulators for \covidn. Error bars indicate 20-80 quantiles for ten random restarts.}
    \label{fig:pareto}
\end{figure*}

\subsection{Simulators}

The effect of a testing policy using the proposed method is tested on two simulators. These simulators are both calibrated to real-world data and account for different contact patterns based on age, profession, and type of household. 

The \openabm\ simulator~\cite{openabm} uses a network-based process to generate contacts, and is calibrated against the UK for different age, household, and occupational networks patterns (school, work, and social network). The simulator has about 150 modifiable parameters, and we use the recommended settings -- the same as used in~\citet{baker2021epidemic,romijnders2022notimetowaste}.  \looseness=-1

The \covasim\ simulator~\cite{covasim} models different contact patterns in layers like households, schools, workplaces, and social communities. Results from this simulator were already used by policymakers~\cite{panovska2020determining,kerr2021controlling}. The contacts are calibrated against a typical city in the USA, and the disease dynamics are stratified for ten age categories. 

\subsection{Experimental Details}\label{sec:experimental_details}

The experiments aim to compare the influence on the peak infection rate for the methods outlined before. Peak infection rate is a common metric to assess the capability of a protocol to mitigate the pandemic~\cite{baker2021epidemic,romijnders2022notimetowaste}. The peak of the pandemic corresponds to most economic and societal consequences, as during that period the hospitals could overfill and governments might decide on a lockdown~\cite {econ_impact_02}. 

\paragraph{Test protocol. }The test protocol is the same in all experiments. Each day, the decentralized algorithm predicts a \covidscore\ per user, and users with the highest score, not currently in quarantine, receive a request to test for \covidn. Simulations on \openabm\ test 10\% of the population daily, and simulations on \covasim\ test only 2\% of the population daily. Positively tested users go in isolation for ten days. We assume a 100\% follow-up from users that are requested to test, Appendix~\ref{app:more_results} explores a scenario where the follow-up is less than 100\%. 

The simulation becomes increasingly challenging when tests for \covidn\ have false positives and false negatives. The default FPR and the FNR are 1\% and 0.1\% respectively. To test the robustness to noisy tests, we increase these noise rates in an experiment similar to~\cite{romijnders2022notimetowaste}.  The FPR increases up to a level of 25\% and the FNR up to a level of 3\% -- these are the worst-case design specifications as prescribed by the European centre for disease control during the COVID19 pandemic~\cite{ecdc}.

\paragraph{Simulation scale. }Unless otherwise noted, we simulate a population of 100.000 users for 100 days for \openabm\ and 91 days in the case of \covasim. At the start of the simulation, 25 people are infected, and interventions start after the third day of the simulation. Whenever a figure depicts an error bar or a table mentions an interval as a subscript, the number indicates the median, and the caps indicate the 20-80 quantiles of ten random restarts. The randomness between different seeds stems from the population dynamics, disease dynamics, and releasing noisy tests for the virus. The unit \permil\ indicates one-per-thousand users.

\paragraph{Differential privacy levels. }The $\delta$ forms an important parameter in differential privacy as this constitutes the probability of exceeding the $\varepsilon$ bound. We set this value to $\frac{1}{1000}$ in all experiments. Existing literature prescribes that the $\delta$ parameter should be smaller than one divided by the data set size~\cite{blanco2022critical, choosing_epsilon, van2023considerations}. Algorithm~\ref{alg:dpfn} uses a privacy bound for the contacts per day, and as our simulators have a max of 200 contacts per day, and an average of only fifteen contacts per day, $\frac{1}{1000}$ is well below the recommended standard.

\subsection{Results for Differential Privacy}
\begin{table}[t]
    \centering
\begin{tabularx}{.9\linewidth}{X|l|l|l}
 \multirow{2}{*}{ \textbf{Test setup}} & \textbf{No} & \multirow{2}{*}{\textbf{DPFN}} & \multirow{2}{*}{\textbf{DPFN+}}\\
  & \textbf{\ \ privacy}  &  &  \rule{0pt}{2.0ex}\\[1mm]
  \hline
(fpr $0.0$\%;   & $0.5$ & $ 1.1$ & $ 0.6$  \rule{0pt}{2.0ex}\\
\ \ \ fnr $0.0$\%) & $\quad ^{[ 0.5,  0.6]}$ & $\quad ^{[ 0.7,  1.4]}$ & $\quad ^{[ 0.5,  0.7]}$ \\[1mm]
(fpr $1$\%;     & $0.5$ & $ 1.1$ & $ 0.6$  \rule{0pt}{2.0ex}\\
\ \ \ fnr $0.1$\%) & $\quad ^{[ 0.4,  0.6]}$ & $\quad ^{[0.9, 1.7]}$ & $\quad ^{[ 0.5,  0.6]}$ \\[1mm]
(fpr $10$\%;    & $0.6$ & $17.6$ & $ 0.9$  \rule{0pt}{2.0ex}\\
\ \ \ fnr $1$\%)   & $\quad ^{[ 0.5,  0.8]}$ & $\quad ^{[11.4, 20.4]}$ & $\quad ^{[ 0.7,  1.0]}$ \\[1mm]
(fpr $25$\%;    & $0.6$ & $46.6$ & $ 0.7$  \rule{0pt}{2.0ex}\\
\ \ \ fnr $3$\%)   & $\quad ^{[ 0.5,  0.8]}$ & $\quad ^{[40.4, 48.0]}$ & $\quad ^{[ 0.6,  0.8]}$ \\[1mm]
\hline   
No testing  & \multicolumn{3}{c}{ $200_{ \ \  [ 190,  212]}$ }  \rule{0pt}{2.0ex}\\[1mm]
\end{tabularx}
    \caption{DP makes the model less robust against noisy tests (column \textit{DPFN}), but more available tests can counteract this effect (column \textit{DPFN+}). This result presents an important message to policymakers. The PIR can be low under private scenarios, but this requires more tests. All results are in 1 daily infection per thousand users (\permil).}
    \label{tab:pir_frac_test_noisy}
\end{table}

\paragraph{Increased privacy at higher performance. } Figure~\ref{fig:pareto} displays our main result, which is a trade-off for the privacy level. The x-axis varies the $\varepsilon$ value for the differential privacy per message, as defined in Equation~\ref{eqn:definition_dp}. On the y-axis, we plot the peak infection rate (PIR). A lower PIR is better, as this corresponds to fewer people simultaneously having the infection. Vice versa, a high PIR implies the occurrence of the pandemic with all its potential consequences. 

Figure~\ref{fig:pareto} shows that for the values of $\varepsilon=1$ per message, DPFN results in a PIR below 1\%, whereas other methods, such as Traditional contact tracing, only achieve a low PIR at ten times as large value for $\varepsilon$. From here onwards, we focus on the $\varepsilon=1$ case as many studies advise this setting for differential privacy~\cite{choosing_epsilon,blanco2022critical,dyda2021differential,choosing_epsilon_02}.

On both simulators, one observes that our DPFN method achieves better PIR than the per-message FN method at $\varepsilon=1$. DPFN introduces noise later in the computation, which we hypothesize maintains a better utility. In terms of PIR, Gibbs sampling does worse than all other methods. This may have two reasons: a) under differential privacy, the Gibbs chain does not converge to the correct distribution~\cite{privacyforfree}, or b) the high number of samples necessitates too high differential privacy budget, $(\varepsilon, \delta)$, to have good utility at low budget. The results for the \covasim\ simulator in the right column of Figure~\ref{fig:pareto} generally display more minor differences in PIR between high and low values of $\varepsilon$. This stems from different modeling assumptions between the \covasim\ and the \openabm\ simulator.

\paragraph{Stability of our DPFN algorithm. }
We also explore the relation between more noisy tests for \covidn\ and the noise due to differential privacy. In Table~\ref{tab:pir_frac_test_noisy}, we increase the FPR and FNR, which has been highlighted as a challenging scenario for traditional contact tracing apps~\cite{security03}. 

Differential privacy makes FN less robust to noisy tests, but this can be counteracted by running more tests. Compared to the \textit{no privacy} column, the column \textit{DPFN} of Table~\ref{tab:pir_frac_test_noisy} shows that under differential privacy, at $\varepsilon=1$, the PIR is an order of magnitude higher. At the highest FPR and FNR, the PIR increases from 0.6 \permil\ to 46.6 \permil. However, adding more available tests can counteract this noise. For column \textit{DPFN+}, we increase the daily testing budget from 10\% to 15\%, and the peak infection rates are again similarly low as the \textit{no privacy} column. This shows that using more available tests can compensate for the noise resulting from differential privacy. Table~\ref{tab:pir_frac_test_noisy} provides an important message to policymakers who need to make a trade-off between infection rates and privacy.

\paragraph{Scaling to 1M agents. } 
Table~\ref{tab:scale} shows the simulation running with $\varepsilon=1$ at different population scales. All simulations in this paper are run with 100.000 users, but this table shows that the benefits of DPFN continue even at the million scale. From as small as fifty thousand users to as large as one million users, the DPFN method results in a significantly lower peak infection rate compared to traditional contact tracing. We emphasize that this is the largest simulation to date for statistical contact tracing.

\begin{table}[t]
    \centering
    \setlength{\tabcolsep}{4pt}
    \begin{tabularx}{\linewidth}{r|l|l}
      \textbf{\#Agents }& \textbf{Traditional} (\permil) & \textbf{DPFN} (\permil)\\[1mm]
      \hline \rule{0pt}{2.0ex}
    $50{,}000$   & $123.2_{ \ \ [108.3,136.4]}$ & $2.5_{ \ \ [1.8,3.7]}$ \\
   $100{,}000$   & $121.6_{ \ \ [110.0,134.3]}$ & $1.5_{\ \ [1.0,1.9]}$ \\
   $500{,}000$   & $134.9_{ \ \ [134.2,135.6]}$ & $0.5_{ \ \ [0.4,0.5]}$ \\
  ${1{,}000{,}000}$   & $134.4_{\ \ [133.8,135.0]}$ & $0.2_{ \ \ [0.2,0.2]}$ \\
    \end{tabularx}
    \caption{Evaluating our algorithm at larger population scales of the simulator. To date, the result with 1 million users is the largest simulation reported for statistical contact tracing. Even at this scale, we show that DPFN results in significantly lower PIR compared to traditional contact tracing.}
    \label{tab:scale}
\end{table}

\section{Discussion and Conclusion}\label{sec:conclusion}

We propose a differentially private algorithm for releasing a \covidscore\ that depends on decentralized communication between contacts. This algorithm protects against a newly identified privacy attack where an adversary aims to reconstruct the \covidscore. Our algorithm results in a two to ten-fold decrease in the peak infection rate compared to other approaches like Gibbs sampling and traditional contact tracing. This improvement holds at $\varepsilon=1$ per message, while other methods only achieve similar results with $\varepsilon \geq 10$. We evaluate the algorithm on two widely used simulators, and we are the first to evaluate these algorithms at a scale of a million agents, where DPFN again achieves lower PIR than traditional contact tracing. 

We see two important directions for future research. First is the study of repeated contacts. Advanced composition bounds are needed to describe privacy when a user has no other contacts but repeated contacts with an adversary. Secondly, our algorithm assumes full adoption of a contact tracing app, but more research is needed into partial adoption. 
We discuss these and other implications of automated decision-making in Appendix~\ref{app:decision}. 

Contact tracing will be one of our first lines of defense to understand and mitigate a virus whenever a new pandemic arises. As argued in the introduction, studies show that privacy concerns are among the top three concerns for adopting a contact tracing app. We believe differential privacy is an essential assurance towards the safe use of contact tracing.

\section*{Acknowledgements}
This work is financially supported by Qualcomm Technologies Inc., the University of Amsterdam and the allowance Top consortia for Knowledge and Innovation (TKIs) from the Netherlands Ministry of Economic Affairs and Climate Policy. Qualcomm AI research is an initiative of Qualcomm Technologies, Inc. and/or its subsidiaries.

\small
\bibliography{aaai24}

@misc{crisp,
  author    = {Ralf Herbrich and
               Rajeev Rastogi and
               Roland Vollgraf},
  title     = {{CRISP:} {A} Probabilistic Model for Individual-Level {COVID-19} Infection
               Risk Estimation Based on Contact Data},
  eprint="2006.04942",
  archivePrefix="arXiv",
  year      = {2020},
}

@book{koller_pgm,
  author    = {Daphne Koller and
               Nir Friedman},
  title     = {Probabilistic Graphical Models - Principles and Techniques},
  publisher = {{MIT} Press},
  year      = {2009},
  bibsource = {dblp computer science bibliography, https://dblp.org}
}

@book{DBLP:books/cu/BV2014,
  author    = {Stephen P. Boyd and
               Lieven Vandenberghe},
  title     = {Convex Optimization},
  publisher    = {Cambridge University Press},
  year      = {2014},
  bibsource = {dblp computer science bibliography, https://dblp.org}
}

@inproceedings{DBLP:conf/csfw/Mironov17,
  author    = {Ilya Mironov},
  title     = {R{\'{e}}nyi Differential Privacy},
  booktitle    = {{IEEE} Computer Security Foundations Symposium},
  year      = {2017},
  bibsource = {dblp computer science bibliography, https://dblp.org}
}

@article{DBLP:conf/uai/Rosen-ZviJY05,
  author    = {Michal Rosen{-}Zvi and
               Michael I. Jordan and
               Alan L. Yuille},
  title     = {The {DLR} Hierarchy of Approximate Inference},
  journal   = {Uncertainty in Artificial Intelligence, UAI},
  year      = {2005},
  bibsource = {dblp computer science bibliography, https://dblp.org}
}

@article{baker2021epidemic,
  title={Epidemic mitigation by statistical inference from contact tracing data},
  author={Baker, Antoine and Biazzo, Indaco and Braunstein, Alfredo and Catania, Giovanni and Dall’Asta, Luca and Ingrosso, Alessandro and Krzakala, Florent and Mazza, Fabio and M{\'e}zard, Marc and Muntoni, Anna Paola and others},
  journal={Proceedings of the National Academy of Sciences},
  year={2021},
  publisher={National Acad Sciences}
}

@article{romijnders2022notimetowaste,
  title={No time to waste: practical statistical contact tracing with few low-bit messages},
  author={Romijnders, Rob and Asano, Yuki and Louizos, Christos and Welling, Max},
  journal={Artificial Intelligence and Statistics Conference, AISTATS},
  year={2023},
}

@article{dwork_dp,
  author    = {Cynthia Dwork and
               Aaron Roth},
  title     = {The Algorithmic Foundations of Differential Privacy},
  journal   = {Foundations and Trends in Theoretical Computer Science},
  year={2014},
}

@article{heikkila2019differentially,
  title={Differentially private markov chain monte carlo},
  author={Heikkil{\"a}, Mikko and J{\"a}lk{\"o}, Joonas and Dikmen, Onur and Honkela, Antti},
  journal={Advances in Neural Information Processing Systems, NeurIPS},
  year={2019}
}

@inproceedings{privacyforfree,
  title={Privacy for free: Posterior sampling and stochastic gradient monte carlo},
  author={Wang, Yu-Xiang and Fienberg, Stephen and Smola, Alex},
  booktitle={International Conference on Machine Learning, ICML},
  year={2015}
}

@inproceedings{DBLP:conf/uai/FouldsGWC16,
  author       = {James R. Foulds and
                  Joseph Geumlek and
                  Max Welling and
                  Kamalika Chaudhuri},
  title        = {On the Theory and Practice of Privacy-Preserving Bayesian Data Analysis},
  booktitle    = {Uncertainty in Artificial Intelligence, UAI},
  year = 2016,
}

@article{tracing_controversy_sk,
    author = {Park, Sangchul and Choi, Gina Jeehyun and Ko, Haksoo},
    title = "{Information Technology–Based Tracing Strategy in Response to COVID-19 in South Korea—Privacy Controversies}",
    journal = {Journal of the American Medical Association},
    year = {2020},
}

@article{tracing_controversy_mobile,
  title={The use of mobile phone data to inform analysis of COVID-19 pandemic epidemiology},
  author={Grantz, Kyra H and Meredith, Hannah R and Cummings, Derek AT and Metcalf, C Jessica E and Grenfell, Bryan T and Giles, John R and Mehta, Shruti and Solomon, Sunil and Labrique, Alain and Kishore, Nishant and others},
  journal={Nature communications},
  year={2020},
}

@misc{central_controversy_app,
  title={Decentralized privacy-preserving proximity tracing},
  author={Troncoso, Carmela and Payer, Mathias and Hubaux, Jean-Pierre and Salath{\'e}, Marcel and Larus, James and Bugnion, Edouard and Lueks, Wouter and Stadler, Theresa and Pyrgelis, Apostolos and Antonioli, Daniele and others},
  eprint="2005.12273",
  archivePrefix="arXiv",
  year={2020}
}

@misc{raskar2020apps,
  title={Apps gone rogue: Maintaining personal privacy in an epidemic},
  author={Raskar, Ramesh and Schunemann, Isabel and Barbar, Rachel and Vilcans, Kristen and Gray, Jim and Vepakomma, Praneeth and Kapa, Suraj and Nuzzo, Andrea and Gupta, Rajiv and Berke, Alex and others},
  eprint="2003.08567",
  archivePrefix="arXiv",
  year={2020}
}

@article{decent_example_singapore,
  title={BlueTrace: A privacy-preserving protocol for community-driven contact tracing across borders},
  author={Bay, Jason and Kek, Joel and Tan, Alvin and Hau, Chai Sheng and Yongquan, Lai and Tan, Janice and Quy, Tang Anh},
  journal={Government Technology Agency-Singapore, Tech. Rep},
  year={2020}
}

@misc{decent_example_pact,
  title={Pact: Privacy sensitive protocols and mechanisms for mobile contact tracing},
  author={Chan, Justin and Foster, Dean and Gollakota, Shyam and Horvitz, Eric and Jaeger, Joseph and Kakade, Sham and Kohno, Tadayoshi and Langford, John and Larson, Jonathan and Sharma, Puneet and others},
  eprint="2004.03544",
  archivePrefix="arXiv",
  year={2020}
}

@misc{DBLP:journals/corr/abs-2003-11511,
  author    = {Hyunghoon Cho and
               Daphne Ippolito and
               Yun William Yu},
  title     = {Contact Tracing Mobile Apps for {COVID-19:} Privacy Considerations
               and Related Trade-offs},
  eprint="2003.11511",
  archivePrefix="arXiv",
  year      = {2020},
  bibsource = {dblp computer science bibliography, https://dblp.org}
}

@inproceedings{econ_impact_01,
  author       = {Doyoung Kim and
                  Hyangsuk Min and
                  Youngeun Nam and
                  Hwanjun Song and
                  Susik Yoon and
                  Minseok Kim and
                  Jae{-}Gil Lee},
  title        = {COVID-EENet: Predicting Fine-Grained Impact of {COVID-19} on Local
                  Economies},
  booktitle    = {Association for the Advancement of Artificial Intelligence, AAAI},
  year      = {2022},
}

@article{econ_impact_02,
title = {Economic impact of COVID-19 pandemic on healthcare facilities and systems: International perspectives},
journal = {Best Practice and Research Clinical Anaesthesiology},
author={Kaye, Alan D and Okeagu, Chikezie N and Pham, Alex D and Silva, Rayce A and Hurley, Joshua J and Arron, Brett L and Sarfraz, Noeen and Lee, Hong N and Ghali, Ghali E and Gamble, Jack W and others},
year = {2021},
}

@article{societal_impact_01,
title = {Addressing the mental health impact of COVID-19 through population health},
journal = {Clinical Psychology Review},
author={Boden, Matt and Zimmerman, Lindsey and Azevedo, Kathryn J and Ruzek, Josef I and Gala, Sasha and Magid, Hoda S Abdel and Cohen, Nichole and Walser, Robyn and Mahtani, Naina D and Hoggatt, Katherine J and others},
year = {2021},
}

@article{societal_impact_02,
title = {COVID-19 pandemic and mental health consequences: Systematic review of the current evidence},
author={Vindegaard, Nina and Benros, Michael Eriksen},
journal = {Brain, Behavior, and Immunity},
year = {2020},
}

@techreport{berger2020seir,
  title={An seir infectious disease model with testing and conditional quarantine},
  author={Berger, David W and Herkenhoff, Kyle F and Mongey, Simon},
  year={2020},
  institution={National Bureau of Economic Research}
}

@article{dyda2021differential,
  title={Differential privacy for public health data: An innovative tool to optimize information sharing while protecting data confidentiality},
  author={Dyda, Amalie and Purcell, Michael and Curtis, Stephanie and Field, Emma and Pillai, Priyanka and Ricardo, Kieran and Weng, Haotian and Moore, Jessica C and Hewett, Michael and Williams, Graham and others},
  journal={Patterns},
  year={2021},
  publisher={Elsevier}
}

@article{li2021impact,
  title={Impact of presymptomatic transmission on epidemic spreading in contact networks: A dynamic message-passing analysis},
  author={Li, Bo and Saad, David},
  journal={Physical Review E},
  year={2021},
  publisher={APS}
}

@misc{covi,
  author    = {Hannah Alsdurf and
               Yoshua Bengio and
               Tristan Deleu and
               Prateek Gupta and
               Daphne Ippolito and
               Richard Janda and
               Max Jarvie and
               Tyler Kolody and
               Sekoul Krastev and
               Tegan Maharaj and
               Robert Obryk and
               Dan Pilat and
               Valerie Pisano and
               Benjamin Prud'homme and
               Meng Qu and
               Nasim Rahaman and
               Irina Rish and
               Jean{-}Franois Rousseau and
               Abhinav Sharma and
               Brooke Struck and
               Jian Tang and
               Martin Weiss and
               Yun William Yu},
  title     = {{COVI} White Paper},
  eprint="2005.08502",
  archivePrefix="arXiv",
  year      = {2020},
  bibsource = {dblp computer science bibliography, https://dblp.org}
}

@article{apple_google_app,
title = {Privacy-preserving contact tracing},
author={Apple and Google},
journal = {apple.com/covid19/contacttracing/, (last accessed August 2023)},
year = {2020},
}

@article{zou2015belief,
  title={A belief propagation approach to privacy-preserving item-based collaborative filtering},
  journal={IEEE Journal of Selected Topics in Signal Processing},
  author={Zou, Jun and Fekri, Faramarz},
  year={2015},
  publisher={IEEE}
}

@article{yildirim2019exact,
  title={Exact MCMC with differentially private moves: revisiting the penalty algorithm in a data privacy framework},
  author={Y{\i}ld{\i}r{\i}m, Sinan and Ermi{\c{s}}, Beyza},
  journal={Statistics and Computing},
  year={2019},
  publisher={Springer}
}

@article{DBLP:journals/tods/ZhangCPSX17,
  author       = {Jun Zhang and
                  Graham Cormode and
                  Cecilia M. Procopiuc and
                  Divesh Srivastava and
                  Xiaokui Xiao},
  title        = {PrivBayes: Private Data Release via Bayesian Networks},
  journal      = {{ACM} Transactions on Database Systems},
  year={2017},
}

@inproceedings{dinur2003revealing,
  title={Revealing information while preserving privacy},
  author={Dinur, Irit and Nissim, Kobbi},
  booktitle={Proceedings of the twenty-second ACM SIGMOD-SIGACT-SIGART symposium on Principles of database systems},
  year={2003}
}

@article{openabm,
  author    = {Robert Hinch and
               William J. M. Probert and
               Anel Nurtay and
               Michelle Kendall and
               Chris Wymant and
               Matthew Hall and
               Katrina A. Lythgoe and
               Ana Bulas Cruz and
               Lele Zhao and
               Andrea Stewart and
               Luca Ferretti and
               Daniel Montero and
               James Warren and
               Nicole Mather and
               Matthew Abueg and
               Neo Wu and
               Olivier Legat and
               Katie Bentley and
               Thomas Mead and
               Kelvin Van{-}Vuuren and
               Dylan Feldner{-}Busztin and
               Tommaso Ristori and
               Anthony Finkelstein and
               David G. Bonsall and
               Lucie Abeler{-}D{\"{o}}rner and
               Christophe Fraser},
  title     = {OpenABM-Covid19 - An agent-based model for non-pharmaceutical interventions
               against {COVID-19} including contact tracing},
  journal   = {PLoS Computational Biology},
  year      = {2021},
}

@book{DBLP:books/sp/RobertC04,
  author    = {Christian P. Robert and
               George Casella},
  title     = {Monte Carlo Statistical Methods},
  publisher = {Springer},
  year      = {2004},
}

@article{bengio2020need,
  title={The need for privacy with public digital contact tracing during the COVID-19 pandemic},
  author={Bengio, Yoshua and Janda, Richard and Yu, Yun William and Ippolito, Daphne and Jarvie, Max and Pilat, Dan and Struck, Brooke and Krastev, Sekoul and Sharma, Abhinav},
  journal={The Lancet Digital Health},
  year={2020},
  publisher={Elsevier}
}

@inproceedings{choosing_epsilon,
  title={Differential privacy: An economic method for choosing epsilon},
  author={Hsu, Justin and Gaboardi, Marco and Haeberlen, Andreas and Khanna, Sanjeev and Narayan, Arjun and Pierce, Benjamin C and Roth, Aaron},
  booktitle={2014 IEEE 27th Computer Security Foundations Symposium},
  year={2014},
  organization={IEEE}
}

@article{choosing_epsilon_02,
  title={Differential privacy: A primer for a non-technical audience},
  author={Wood, Alexandra and Altman, Micah and Bembenek, Aaron and Bun, Mark and Gaboardi, Marco and Honaker, James and Nissim, Kobbi and O'Brien, David R and Steinke, Thomas and Vadhan, Salil},
  journal={Vanderbilt Journal of Entertainment and Technology Law},
  year={2018},
}

@article{vadrevu2020hybrid,
  title={A hybrid approach for personal differential privacy preservation in homogeneous and heterogeneous health data sharing},
  author={Vadrevu, Pavan Kumar and Adusumalli, Sri Krishna and Mangalapalli, Vamsi Krishna},
  journal={High Technology Letters},
  year={2020}
}

@techreport{vepakomma2021dams,
  title={DAMS: Meta-estimation of private sketch data structures for differentially private COVID-19 contact tracing},
  author={Vepakomma, Praneeth and Pushpita, Subha Nawer and Raskar, Ramesh},
  year={2021},
  institution={Accessed: June 14, 2023. [Online]. Available: https://www.media.mit.edu/publications/dams-meta-estimation-of-private-sketch-data-structures-for-differentially-private-covid-19-contact-tracing/}
}

@article{ecdc,
  author    = {ECDC},
  title     = {Considerations on the use of self-tests for COVID-19 in the EU/EEA},
  journal   = {European Centre for Disease prevention \text{and} control,  technical report, 17 March 2021},
  year      = {2021},
}

@article{kermack1927contribution,
  title={A contribution to the mathematical theory of epidemics},
  author={Kermack, William Ogilvy and McKendrick, Anderson},
  journal={Proceedings of the royal society of London},
  year={1927},
}

@book{anderson1992infectious,
  title={Infectious diseases of humans: dynamics and control},
  author={Anderson, Roy M and May, Robert M},
  year={1992},
  publisher={Oxford university press}
}

@article{covasim,
    author = {Kerr, Cliff C. AND Stuart, Robyn M. AND Mistry, Dina AND Abeysuriya, Romesh G. AND Rosenfeld, Katherine AND Hart, Gregory R. AND Núñez, Rafael C. AND Cohen, Jamie A. AND Selvaraj, Prashanth AND Hagedorn, Brittany AND George, Lauren AND Jastrzebski, Michal AND Izzo, Amanda S. AND Fowler, Greer AND Palmer, Anna AND Delport, Dominic AND Scott, Nick AND Kelly, Sherrie L. AND Bennette, Caroline S. AND Wagner, Bradley G. AND Chang, Stewart T. AND Oron, Assaf P. AND Wenger, Edward A. AND Panovska-Griffiths, Jasmina AND Famulare, Michael AND Klein, Daniel J.},
    journal = {PLOS Computational Biology},
    title = {Covasim: An agent-based model of COVID-19 dynamics and interventions},
    year = {2021},
}

@article{burdinski2022understanding,
  title={Understanding the impact of digital contact tracing during the COVID-19 pandemic},
  author={Burdinski, Angelique and Brockmann, Dirk and Maier, Benjamin Frank},
  journal={PLOS Digital Health},
  year={2022},
  publisher={Public Library of Science San Francisco, CA USA}
}

@article{blanco2022critical,
  title={A critical review on the use (and misuse) of differential privacy in machine learning},
  author={Blanco-Justicia, Alberto and S{\'a}nchez, David and Domingo-Ferrer, Josep and Muralidhar, Krishnamurty},
  journal={ACM Computing Surveys},
  year={2022},
  publisher={ACM New York, NY}
}

@misc{braunstein2023small,
  title={Small-Coupling Dynamic Cavity: a Bayesian mean-field framework for epidemic inference},
  author={Braunstein, Alfredo and Catania, Giovanni and Dall'Asta, Luca and Mariani, Matteo and Mazza, Fabio and Tarabolo, Mattia},
  eprint="2306.03829",
  archivePrefix="arXiv",
  year={2023}
}

@article{security01,
  title={A survey of COVID-19 contact tracing apps},
  author={Ahmed, Nadeem and Michelin, Regio A and Xue, Wanli and Ruj, Sushmita and Malaney, Robert and Kanhere, Salil S and Seneviratne, Aruna and Hu, Wen and Janicke, Helge and Jha, Sanjay K},
  journal={IEEE access},
  year={2020},
}

@inproceedings{security02,
  title={Cross hashing: Anonymizing encounters in decentralised contact tracing protocols},
  author={Ali, Junade and Dyo, Vladimir},
  booktitle={IEEE International Conference on Information Networking },
  year={2021},
}

@article{security03,
  title={A survey of automatic contact tracing approaches using Bluetooth Low Energy},
  author={Reichert, Leonie and Brack, Samuel and Scheuermann, Bj{\"o}rn},
  journal={ACM Transactions on Computing for Healthcare},
  year={2021},
}

@misc{van2023considerations,
  title={Considerations on the Theory of Training Models with Differential Privacy},
  author={van Dijk, Marten and Nguyen, Phuong Ha},
  eprint="2303.04676",
  archivePrefix="arXiv",
  year={2023}
}

@article{perra2021non,
  title={Non-pharmaceutical interventions during the COVID-19 pandemic: A review},
  author={Perra, Nicola},
  journal={Physics Reports},
  year={2021}
}

@inproceedings{abadi2016deep,
  title={Deep learning with differential privacy},
  author={Abadi, Martin and Chu, Andy and Goodfellow, Ian and McMahan, H Brendan and Mironov, Ilya and Talwar, Kunal and Zhang, Li},
  booktitle={ACM SIGSAC conference on computer and communications security},
  year={2016}
}

@article{privacy_concern_01,
  title={To use or not to use a COVID-19 contact tracing app: Mixed methods survey in Wales},
  author={Jones, Kerina and Thompson, Rachel and others},
  journal={JMIR mHealth and uHealth},
  year={2021}
}

@article{privacy_concern_02,
  title={Drivers of downloading and reasons for not downloading COVID-19 contact tracing and exposure notification apps: A national cross-sectional survey},
  author={Gao, Golden and Lang, Raynell and Oxoby, Robert J and Mourali, Mehdi and Sheikh, Hasan and Fullerton, Madison M and Tang, Theresa and Manns, Braden J and Marshall, Deborah A and Hu, Jia and others},
  journal={PLOS one},
  year={2022},
  publisher={Public Library of Science San Francisco, USA}
}

@article{privacy_concerns_03,
  title={Reasons for nonuse, discontinuation of use, and acceptance of additional functionalities of a COVID-19 contact tracing app: cross-sectional survey study},
  author={Walrave, Michel and Waeterloos, Cato and Ponnet, Koen},
  journal={JMIR Public Health and Surveillance},
  year={2022},
  publisher={JMIR Publications Toronto, Canada}
}

@inproceedings{lam2015numba,
  title={Numba: A llvm-based python jit compiler},
  author={Lam, Siu Kwan and Pitrou, Antoine and Seibert, Stanley},
  booktitle={Proceedings of the Second Workshop on the LLVM Compiler Infrastructure in HPC},
  year={2015}
}

@article{gil2013renyi,
  title={R{\'e}nyi divergence measures for commonly used univariate continuous distributions},
  author={Gil, Manuel and Alajaji, Fady and Linder, Tamas},
  journal={Information Sciences},
  year={2013},
  publisher={Elsevier}
}

@inproceedings{user_level_privacy,
  author       = {H. Brendan McMahan and
                  Daniel Ramage and
                  Kunal Talwar and
                  Li Zhang},
  title        = {Learning Differentially Private Recurrent Language Models},
  booktitle    = {International Conference on Learning Representations, ICLR},
  year         = {2018},
}

@article{panovska2020determining,
  title={Determining the optimal strategy for reopening schools, the impact of test and trace interventions, and the risk of occurrence of a second COVID-19 epidemic wave in the UK: a modelling study},
  author={Panovska-Griffiths, Jasmina and Kerr, Cliff C and Stuart, Robyn M and Mistry, Dina and Klein, Daniel J and Viner, Russell M and Bonell, Chris},
  journal={The Lancet Child \& Adolescent Health},
  year={2020},
  publisher={Elsevier}
}

@article{kerr2021controlling,
  title={Controlling COVID-19 via test-trace-quarantine},
  author={Kerr, Cliff C and Mistry, Dina and Stuart, Robyn M and Rosenfeld, Katherine and Hart, Gregory R and N{\'u}{\~n}ez, Rafael C and Cohen, Jamie A and Selvaraj, Prashanth and Abeysuriya, Romesh G and Jastrzebski, Michal and others},
  journal={Nature communications},
  year={2021},
  publisher={Nature Publishing Group}
}

@inproceedings{beaver1990round,
  title={The round complexity of secure protocols},
  author={Beaver, Donald and Micali, Silvio and Rogaway, Phillip},
  booktitle={ACM symposium on Theory of computing},
  year={1990}
}

@inproceedings{ben2016optimizing,
  title={Optimizing semi-honest secure multiparty computation for the internet},
  author={Ben-Efraim, Aner and Lindell, Yehuda and Omri, Eran},
  booktitle={ACM SIGSAC Conference on Computer and Communications Security},
  year={2016}
}

@inproceedings{sabt2015trusted,
  title={Trusted execution environment: what it is, and what it is not},
  author={Sabt, Mohamed and Achemlal, Mohammed and Bouabdallah, Abdelmadjid},
  booktitle={IEEE Trustcom/BigDataSE/Ispa},
  year={2015},
}

@inproceedings{DBLP:conf/icml/KairouzOV15,
  author       = {Peter Kairouz and
                  Sewoong Oh and
                  Pramod Viswanath},
  title        = {The Composition Theorem for Differential Privacy},
  booktitle    = {International Conference on Machine Learning, {ICML}},
  year         = {2015},
}

@inproceedings{farrand2020neither,
  title={Neither private nor fair: Impact of data imbalance on utility and fairness in differential privacy},
  author={Farrand, Tom and Mireshghallah, Fatemehsadat and Singh, Sahib and Trask, Andrew},
  booktitle={Proceedings of the 2020 workshop on privacy-preserving machine learning in practice},
  year={2020}
}

@inproceedings{balle2018gauss,
  title={Improving the Gaussian Mechanism for Differential Privacy: Analytical Calibration and Optimal Denoising},
  author={Balle, Borja and Wang, Yu-Xiang},
  booktitle= {International Conference on Machine Learning, {ICML}},
  year={2018}
}
\normalsize

\appendix

\iftrue
\cleardoublepage

\section{Additional Results} \label{app:more_results}

\begin{figure*}[t]
    \centering
    \includegraphics[width=.95\textwidth]{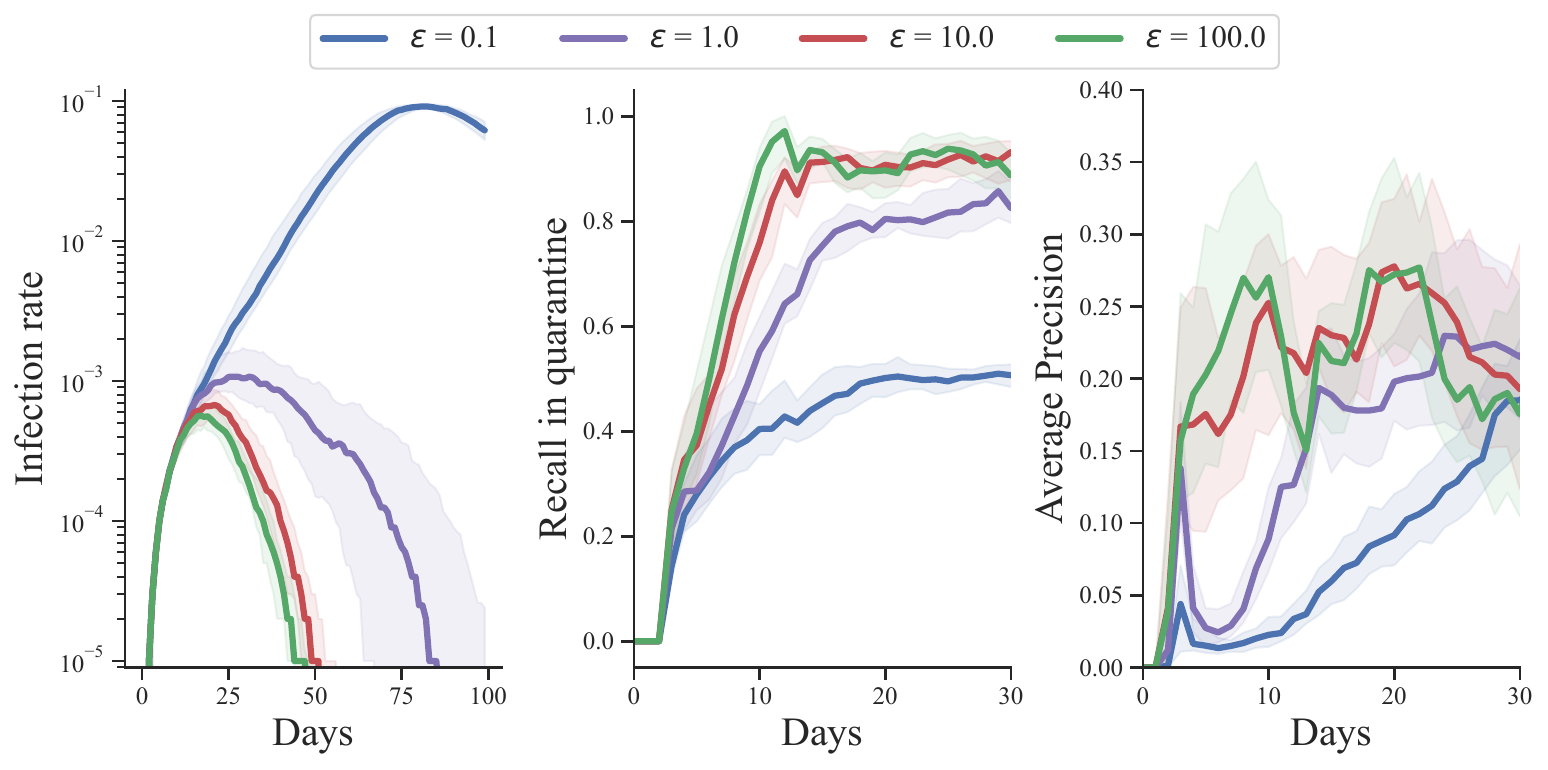}
    \caption{ Recall and average precision during a simulation on \openabm. The shaded regions indicate the 20-80 quantiles of twenty random restarts. The peak infection rate of the curve $\varepsilon=1$ is in between lower and higher privacy levels. This effect is observed as well in intermediate values for the recall and average precision during the crucial first month of the epidemic simulation. The recall and average precision diagrams only plot 30 days of simulation, which is the crucial phase for a pandemic~\cite{perra2021non}.}
    \label{fig:recalls}
\end{figure*}

Our paper presents two trade-offs when designing a DP contact tracing algorithm. Figure~\ref{fig:pareto} presents a trade-off with the level of privacy. In this extra analysis, we look further at the financial cost of contact tracing. A contact tracing algorithm, fundamentally, decides who to get tested based on earlier test results and the contacts between users. With fewer tests daily available, the \covidscore\ is less precise and achieving a low PIR becomes harder. Producing extra available tests comes at a cost, and restrictions may apply. To this end, we vary the number of available tests while setting DP at the conventionally accepted value of $\varepsilon=1$, and run this simulation on the \covasim\ simulator.  Figure~\ref{fig:fractiontest} shows the result of this experiment. For a small number of tests, both DPFN and traditional contact tracing result in a too high peak infection rate. Achieving a result below 0.003 PIR, the traditional contact tracing requires $10{,}000$ tests, while DPFN requires only $500$ tests. We conclude from this figure that DPFN achieves a low PIR with fewer tests. Additionally, this figure shows the trade-off between peak infection rate and financial cost in the form of the number of available tests.  \\ 

This section further presents three additional results.

\textbf{Plotting recall and average precision: } Figure~\ref{fig:recalls} uses the same experimental settings as Figure~\ref{fig:pareto} and plots the individual timesteps on the x-axis. The shaded region indicates the 20-80 quantiles of 20 random restarts. Recall is calculated as the ratio of the number of infectious agents in quarantine divided by the total number of infectious agents. Average precision is calculated on the \covidscore\ against the binary indicator of being infectious -- a high average precision means the \covidscore\ can be used to discriminate between infectious and non-infectious users.

The Figure plots results for high $\varepsilon$ values (10, 100) for comparison. These values are not considered private enough~\cite{choosing_epsilon,choosing_epsilon_02}, but show a comparison for recall and average precision. Conversely, setting $\varepsilon=0.1$ results in too high infection rates. The curve of $\varepsilon=1$ strikes a balance between differential privacy and low infection rates. 

\textbf{Scale experiment on \covasim\ simulator: } We reproduce the scale experiments of Table~\ref{tab:scale}. Table~\ref{tab:scale_cv} shows the result. Due to the differences in dynamics across different scales, the peak infection rates at particular scales vary. However, our DPFN method achieves a lower peak infection rate across different simulation scales.

\textbf{Loss to follow-up: } Here we vary the loss to follow-up from a test result. All experiments assume a user isolates after receiving a positive test for \covidn. If the user ignores this signal, there are two consequences: a) a user that is likely infectious continues to interact with other users, possibly spreading the virus; b) there is an opportunity cost as the test might have been sent to another user. The \covasim\ simulator can simulate this scenario where a user fails to follow up on a positive test. We directly modulate the \texttt{loss-prob} parameter in their source code, and Figure~\ref{fig:loss_follow_up} shows the results. One can make two observations from the figure. Firstly, up to a follow-up loss of 75\%, the DPFN algorithm achieves lower PIR than the traditional method. Secondly, the DPFN method shows robustness by getting lower than 0.5\% PIR for a probability of loss up to 20\%.

\begin{figure}
    \centering
    \includegraphics[width=\linewidth]{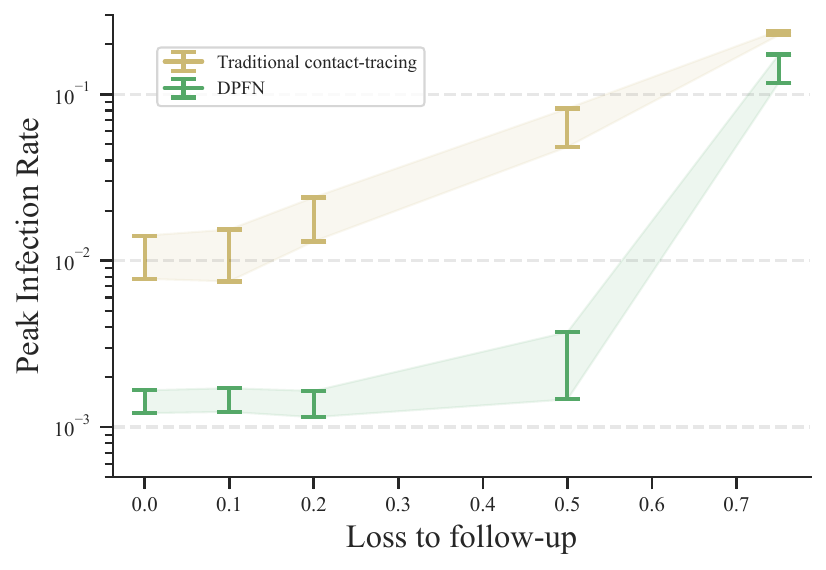}
    \caption{This figure explores the scenario where a user tests positive, but does not isolate. `Loss to follow-up' is the probability that a user ignores the request to isolate after a positive test. The DPFN method achieves lower PIR than traditional contact tracing across a wide range of the `loss to follow-up' probability.}
    \label{fig:loss_follow_up}
\end{figure}

\begin{figure}
    \centering
    \includegraphics[width=\linewidth]{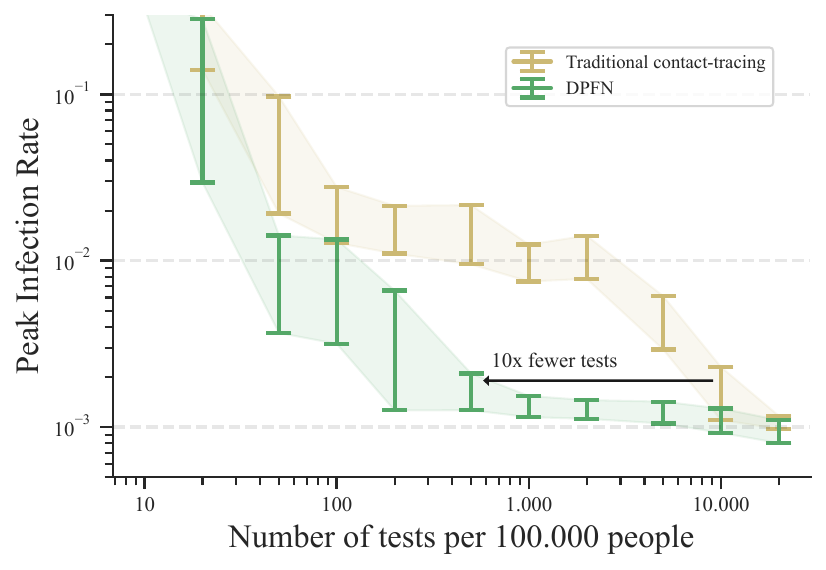}
    \caption{We vary the number of tests daily available in the simulation. Generally, more available tests provide more information to the contact tracing algorithm -- c.f. Table~\ref{tab:pir_frac_test_noisy}. Considering a peak infection rate of 0.003, the traditional contact tracing method achieves this low PIR with $10{,}000$ tests, while DPFN only requires $500$ tests for the same PIR. This figure is made with $\varepsilon=1$. \looseness=-1}
    \label{fig:fractiontest}
\end{figure}

\section{Algorithmic Decision-Making}\label{app:decision}

This section provides background on three important topics in this paper.

\textbf{Differential privacy.} We use differential privacy to quantify the level of privacy when releasing a single covidscore. Besides the conventional definition in Equation~\ref{eqn:definition_dp}, there is a more intuitive interpretation in terms of type 1 and type 2 error rates that an adversary could have if an attack is attempted~\cite{DBLP:conf/icml/KairouzOV15}. The DP noise inherently decreases the algorithm's utility. To this end, we have plotted the trade-offs of infection rate against privacy level in Figure~\ref{fig:pareto} and against financial cost in the form of testing capacity in Figure~\ref{fig:fractiontest}. We hope these figures can inform policymakers about the available settings for differentially private contact tracing algorithms.

\textbf{Unintended consequences and use of data.} All algorithms in this paper use only contacts and \covidn\ tests as input data. No other features about users are used. We assume that contacts have been established securely, for which we cite references in Section~\ref{sec:related_work}. We assume that the test results are processed in a trusted execution environment~\cite{sabt2015trusted}. 

Despite these restrictions on the dataset, there are open questions about the consequences of automating societal decisions. In our particular setting, we see at least three relevant issues.
Firstly, unequal access to smartphones could have unintended consequences, such as disparate infection rates in particular groups with unequal access to smartphones. Secondly, even under equal access, partial app adoption could reduce performance due to the lack of information and whether a contact occurred. We studied a somewhat related effect in Figure~\ref{fig:loss_follow_up}, but more research on this topic is necessary. Lastly, previous research has shown that DP can exacerbate biases existing in the data. This effect has been studied in~\citet{farrand2020neither} and should be investigated in the decentralized setting of contact tracing. We believe that studying these effects holistically is important before deploying algorithmic decision-making in the real world.

\textbf{Data retention.} The data retention of a potential app should be limited in scope. To this end, all algorithms in this paper use the data for only a fourteen-day window, and the data is deleted automatically afterward. Our implementation of DPFN is purely stateless. This means no state about a person or health status is maintained; data can be removed at any time during the time window, and removed contacts will not influence any prediction being made the next day. Appendix Section~\ref{app:experimental_details} and Table~\ref{tab:hyper_numdays} report on the impact of setting such a window and show that any window length longer than fourteen days only slightly improves the results.

\section{Additional Details on the Method}

\subsection{Notation}

In this paper and the rest of the appendix, we use the following notation:

\begin{itemize}
    \item $z_{u,t} \in \{S, E, I, R\}$: The random variable which takes on value for the states Susceptible, Exposed, Infectious, Recovered;
    \item $N(u,t)$: the set of neighbors of user $u$ at timestep $t$. When it is clear from context $N(u)$ or $N(t)$ might be used to abbreviate for $N(u,t)$;
    \item $z_{N(u,t)}$: the set of random variables that are in the neighborhood of $z_{u,t}$, these are the states of contacts of user $u$ at time step $t$;
    \item $D_\mathcal{O} = \{o_{u_i,t_i}  \}_{i=1}^O$: The data set of observations with $O$ observations, each with an outcome $\{0,1\}$ for user $u_i$ at time step $t_i$. These observations have false positives and false negatives with respect to the underlying state; this observation model is mentioned in Equation~\ref{eqn:test_outcome}.
    \item $\phi_{c,t} \in [0,1)$: General \covidscore\ for a contact $c$ that occurred on time $t$;
    \item $\omega_{c,t} \in (1-p_1, 1]$: Message from a contact $c$ at time step $t$. Related to $\phi_{c,t}$ by Equation~\ref{eqn:phi_to_S};
    \item $C \in \mathbb{N}$: capital letter $C$ is generally used to indicate the number of contacts, $C_t = |N(t)|$ for a particular user;
    \item $p_0, p_1 \in [0, 1)$: model parameters, $p_0$ is the probability for spontaneous infection, $p_1$ is the probability of the transmission of the virus given the occurrence of a contact and is used in Equation~\ref{eqn:crisp_noisy_or};
    \item $g, h \in (0, 1)$: model parameters, $g$ is the probability of the state transition $E \rightarrow I$, $h$ is the probability of the state transition $I \rightarrow R$;
    \item $b_u(z_u)$: FN belief over random variable $z_u$;
    \item $B_{N(u)}$: joint beliefs over a set of random variables, in this case, $N(u)$ is the set of neighbors, contacts, of user $u$;
    \item $\mu, \sigma^2 \in \mathbb{R}^+$: general parameters for the $\operatorname{log-normal}$ distribution;
    \item $a > 1, \rho > 0$: $a$ is the order of the R\'enyi divergence, $\rho$ is the upper bound on the R\'enyi divergence;
    \item $D_a(\cdot | \cdot)$: the divergence of order $a$ between two probability distributions, in this paper we solely use R\'enyi divergence;
    \item $\varepsilon > 0, \delta \in [0,1]$: parameters for $(\varepsilon,\delta)$ Differential privacy, defined in Equation~\ref{eqn:definition_dp}; for brevity, normal letter $d$ refers to $\log\frac{1}{\delta}$;
    \item $F(\cdot), F_1(\cdot), F_2(\cdot)$: these functions all refer to the Factorized Neighbor function, this approximate inference method was introduced by~\cite{DBLP:conf/uai/Rosen-ZviJY05} and the update equations for an SEIR epidemic model were derived by~\cite{romijnders2022notimetowaste};
    \item $\alpha, \beta$: False Positive and False negative rate for the \covidn\ tests, these parameters are assumed to be known by the statistical model during inference;
    \item $\gamma$ generally indicates a clipping value. For the per-message methods, values are clipped in $[\gamma, 1-\gamma]$, for the DPFN, messages are clipped in $[\gamma_l, \gamma_u]$
\end{itemize}

\subsection{Properties of the Log-Normal Distribution}\label{app:lognormal_properties}

Our method makes extensive use of the log-normal distribution and we outline a few essential properties in this section. We define as log-normal distribution:

\begin{align}
    p(x) &= \frac{1}{x \sigma \sqrt{2 \pi}} \quad \exp \{ \frac{-1}{2 \sigma^2} \left( \log(x) - \mu \right)^2 \}  
\end{align}    

This distribution has mean parameter $\mu \in \mathbb{R}$, and variance parameter $\sigma \in \mathbb{R}^+$. The domain is $x \in \mathbb{R}^+$, and the expected value is

\begin{equation} \label{eqn:log-normal_mu}
    m = E[x] = \exp \{ \mu + \frac{\sigma^2}{2} \}. 
\end{equation}

We will refer to $\mu$ as the mean parameter and refer to $m$ as the expected value. Similarly, we will refer to $\sigma^2$ as the variance parameter.

One can rewrite Equation~\ref{eqn:log-normal_mu} for the $\mu$ parameter:
\begin{align}
    \mu &=  \log(m) - \frac{\sigma^2}{2}  \label{eqn:mu_log-normal_bias_correction}.
\end{align}

The R\'enyi divergence between two probability distributions, $p_u$ and $p_v$, and respective parameters, $\mu_u$, $\sigma_u^2$, $\mu_v$, $\sigma^2_v$~\cite{gil2013renyi} is the following:

\begin{align}
& D_{a}(p_u \vert p_v) = \nonumber \\
    & \log \left( \frac{\sigma_v}{\sigma_u} \right)+ \frac{1}{2(a - 1)}\log \left( \frac{\sigma_v^2}{\sigma_*^2} \right) + \frac{a}{2} \frac{(\mu_u - \mu_v)^2}{\sigma_*^2} \label{eqn:renyi_divergence}\\ 
    & \nonumber \\
    & \sigma_*^2 = a \sigma^2_v + (1-a) \sigma^2_u. 
\end{align}

The family of log-normal distributions is closed under multiplication. If $X_{j}\sim \operatorname {log-normal} (\mu _{j},\sigma _{j}^{2})$ are n independent, log-normally distributed variables, then:

\begin{equation} 
Y=\textstyle \prod _{j=1}^{n}X_{j}\sim \operatorname{log-normal} {\Big (}\textstyle \sum _{j=1}^{n}\mu _{j},\ \sum _{j=1}^{n}\sigma _{j}^{2}{\Big )}. 
\label{eqn:log-normal_addition}
\end{equation}

\subsection{R\'enyi Differential Privacy of Algorithm~\ref{alg:dpfn}}\label{app:detailed_RDP}

This section derives the R\'enyi differential privacy for the method outlined in Algorithm~\ref{alg:dpfn}. Previous work established a relation between bounding the R\'enyi divergence and converting the bound to the $\varepsilon$ and $\delta$ for DP~\cite{DBLP:conf/csfw/Mironov17}. As such, we bound the R\'enyi divergence between the product terms that are required each day for the computation of $F_2(\cdot)$ (defined in Section~\ref{sec:method}). The conversion from $(a,\rho)$-RDP to $(\varepsilon,\delta)$ is discussed in Appendix~\ref{app:a_opt}.

Appendix~\ref{app:lognormal_properties} overviews properties of the log-normal distribution. We know that the log-normal distribution is closed under multiplication, and its R\'enyi divergence has a closed-form expression~\cite{gil2013renyi}. We will bound the R\'enyi divergence of two product terms, $\omega_{*,t}$,  under the worst-case adjacent data sets. In Section~\ref{sec:method} we showed that FN computation depends on a product of messages $\omega_{c,t}$. The value of $\omega_{u,t}$ represents a function of the \covidscore\ that a contact communicates by the relation $\omega_{c, t} = 1 - p_1 \phi_{c,t}$. This is explained after Equation~\ref{eqn:phi_to_S}. 

We start by writing the R\'enyi divergence of this product term. Take two adjacent data sets, according to Equation~\ref{eqn:defn_neighbor}, and name the distributions over their product $p_u$ and $p_v$, the R\'enyi divergence is:

\begin{align}
    D_{a}(p_u \vert p_v) &= \underbrace{\log \left( \frac{\sigma_v}{\sigma_u} \right)+ \frac{1}{2(a - 1)}\log \left( \frac{\sigma_v^2}{\sigma_*^2} \right)}_{\text{equals 0}} \nonumber \\ 
    & \qquad + \frac{a}{2\sigma_*^2 } \cdot (\mu_u - \mu_v)^2 \label{eqn:renyi_bound_zero_terms} \\
     &= \frac{a}{2\sigma_*^2 } (\mu_u - \mu_v)^2  = \frac{a}{2 C  \sigma^2 } (\mu_u - \mu_v)^2 \label{eqn:div_bound_01_app}
\end{align}

The first two terms in Equation~\ref{eqn:renyi_bound_zero_terms} are zero as each dataset for the same number of contacts has the same variance parameter $\sigma^2_* =\sigma^2_v = \sigma_u^2 =  C \cdot \sigma^2$. The variance parameter for a single $\omega_{c,t}$-term is written as $\sigma^2$. As per the property of the log-normal distribution, the $\sigma^2$ parameters are additive when the random variable is multiplied, and $C \cdot \sigma^2$ is the variance parameter of the product.

\begin{table}[t]
    \centering
    \setlength{\tabcolsep}{4pt}
    \begin{tabularx}{\linewidth}{r|l|l}
      \textbf{\#Agents } & \textbf{Traditional} (\permil) & \textbf{FN} (\permil)\\[1mm]
      \hline \rule{0pt}{2.0ex}
    $50{,}000$       &  $7.6_{ \ \ [5.8,9.9]}$  & $1.1_{ \ \ [0.8, 1.4]}$ \\
   $100{,}000$       & $11.6_{ \ \ [7.8,14.1]}$ & $1.0_{ \ \  [1.0, 1.4]}$ \\
   $500{,}000$       &  $2.1_{ \ \ [0.5,2.7] }$ & $0.2_{ \ \ [0.2, 0.3]}$ \\
  ${1{,}000{,}000}$  &  $1.9_{ \ \ [1.0,2.6]}$  & $0.2_{ \ \ [0.2, 0.2]}$ \\
    \end{tabularx}
    \caption{Reproducing Table~\ref{tab:scale} on the \covasim\ simulator. These simulators have different dynamics. Still, across scales, our differentially private decentralized algorithm achieves lower peak infection rates. }
    \label{tab:scale_cv}
\end{table}

The R\'enyi divergence depends on the difference in the mean parameters, $(\mu_u - \mu_v)^2$, and we want to bound this for the worst-case adjacent data sets. A log-normal distribution has mean $m = \operatorname{exp}\{\mu + \frac{\sigma^2}{2}\}$. As such, we define the $\mu$ parameter of the log-normal distribution as:

\begin{equation} \label{eqn:rdp_log_mean}
\mu_{c,t} = \log \omega_{c,t} - \frac{\sigma^2}{2}.
\end{equation}

Each message is sampled from a log-normal distribution like $s_{c,t} \sim \operatorname{log-normal}(\mu_{c,t}, \sigma^2)$, with $\mu_{i, t} = \log \omega_{i, t} - \frac{\sigma^2}{2}$. The product is $\omega_{*, t} = \prod_{i=1}^C \omega_{i, t}$, which has a distribution $s_{*, t} \sim \operatorname{log-normal}(\mu=\sum_{i=1}^C \mu_{c,t}, C\cdot\sigma^2)$.

The $\mu$ parameter of a single message is defined as Equation~\ref{eqn:rdp_log_mean}. Therefore, the $\mu$ parameter of the product follows from:

\begin{align}
    \mu_{*,t} &= \sum_{i=1}^C \mu_{c,t} = (\sum_{c \in N(u,t)} \log [ \omega_{c,t}] )- \frac{C\sigma^2}{2} \label{eqn:rdp_detailed_sum}
\end{align}

Using FN for decentralized contact tracing, we can clip the \covidscore\ $\phi_{c,t}$ in the range $[\gamma_l, \gamma_u]$. If so, the messages $\omega_{c,t}$ will be bounded as well between $[1-p_1\gamma_u, 1-p_1\gamma_l]$. \looseness=-1

The sum in Equation~\ref{eqn:rdp_detailed_sum} is monotone in each message $\omega_{c,t}$. We aim to bound the worst-case value for $\mu_u - \mu_v$ and thus we have two cases. Firstly, each message is at the lower clip value $\gamma_l$, against the adjacent data set where one message is at the upper clip value $\gamma_u$. Secondly, each message is at the upper clip value $\gamma_u$, against the adjacent data set where one message is at the lower clip value $\gamma_l$. 

In the first case, we have
\begin{align}
        & (\mu_u - \mu_v) = \underbrace{C (\log(1 - \gamma_l p_1 ) - \frac{\sigma^2}{2})}_{\text{all $\gamma_l$}} \nonumber \\
        & - \left( \underbrace{(C-1)(\log(1 - \gamma_l p_1 ) - \frac{\sigma^2}{2}) + \log(1 - \gamma_u p_1 ) - \frac{\sigma^2}{2}}_{\text{one switch to $\gamma_u$}}  \right) \\
        &= \log(1 - \gamma_l p_1) - \log(1 - \gamma_u p_1) \nonumber.
    \end{align}

In the second case, we have
\begin{align}
        & (\mu_u - \mu_v) = \underbrace{C (\log(1 - \gamma_u p_1 ) - \frac{\sigma^2}{2})}_{\text{all $\gamma_u$}} - \nonumber \\ 
        & \left( \underbrace{(C-1)(\log(1 - \gamma_u p_1 ) - \frac{\sigma^2}{2}) + \log(1 - \gamma_l p_1 ) - \frac{\sigma^2}{2}}_{\text{one switch to $\gamma_l$}}  \right) \\ 
        &= \log(1 - \gamma_u p_1) - \log(1 - \gamma_l p_1) \nonumber.
\end{align}

After squaring, we have in both cases
\begin{equation}
    (\mu_u - \mu_v)^2 = (\log(1 - \gamma_u p_1) - \log(1 - \gamma_l p_1))^2 \label{eqn:rdp_worst_case_mus}.
\end{equation}

Equation~\ref{eqn:rdp_worst_case_mus} is established for the worst-case adjacent data sets. Therefore, we have that for any adjacent data sets
 \begin{align}
    & D_{a}(p_u \vert p_v) \leq  \frac{a}{2C  \sigma^2  }(\log(1 - \gamma_u p_1) - \log(1 - \gamma_l p_1))^2  = \rho \nonumber \\ 
    & \quad \text{for any two adjacent data sets} \label{eqn:rdp_log-normal}.
\end{align}

In Equation~\ref{eqn:rdp_log-normal}, we write the upper bound as $\rho$. Therefore, we have $(a,\rho)$-RDP whenever $\sigma^2$ has at least the following value
\begin{align}
    \sigma^2 \geq \frac{a}{2C  \rho  }(\log(1 - \gamma_u p_1) - \log(1 - \gamma_l p_1))^2 \label{eqn:rdp_scales_a_e}
\end{align}

\textit{Difference from Gaussian mechanism in the log-domain: } Other than RDP with a log-normal distribution, one could also use the Gaussian mechanism from~\cite{dwork_dp,balle2018gauss} in the log-domain. Such a method would apply the Gaussian mechanism to the logarithm of each score and consider the exponent function as post-processing. We experimented with this method and found this method to achieve strictly worse PIR results. One reason could be the form of Equation~\ref{eqn:rdp_log_mean}. Due to Jensen's inequality, we know that the mean of a concave function (e.g. the logarithm) is smaller than or equal to the concave function applied to the mean. The factor $-\frac{\sigma^2}{2}$ could be considered to counteract this `bias,' and we hypothesize that this is the reason the Gaussian mechanism in the log domain achieves worse results than RDP with the log-normal distribution. 

\textit{Clipping with public knowledge: } The log-normal distribution assumes a value on $\mathbb{R}^+$. In the product of Equation~\ref{eqn:fn_needs_only_product}, $C$ \covidscore\ are clipped to $[\gamma_l, \gamma_u]$, and each message is calculated according to Equation~\ref{eqn:phi_to_S}. As such, it is public knowledge that the product of messages should lie in $[(1-\gamma_u p_1)^C, (1-\gamma_l p_1)^C]$. Therefore, after the sampling in line 5 of Algorithm~\ref{alg:dpfn}, we clip the messages to this known interval. \looseness=-1 

\subsection{Optimize for Parameters of RDP} \label{app:a_opt}

Section~\ref{sec:dp_log_normal} presents an algorithm that achieves differential privacy for contact tracing via R\'enyi differential privacy. Although R\'enyi differential privacy has better composition properties~\cite{DBLP:conf/csfw/Mironov17}, the analysis introduces a new hyperparameter $a$. This hyperparameter could be optimized via an experimental parameter sweep~\cite{abadi2016deep}. Fortunately, for our particular problem, we find a closed-form solution for this hyperparameter. This allows one to convert $(a,\rho)$-RDP to $(\varepsilon,\delta)$-DP. In this section, we write out the system of equations and show four closed-form solutions to the resulting polynomial. 

We reduce the equations in Section~\ref{sec:dp_log_normal} to an optimization problem and use the KKT conditions to find a stationary point. Equation~\ref{eqn:rdp_scales_a_e} shows that the noise scale grows linearly with $\frac{a}{\rho}$. Arguably, any lower noise scale implies less noise on the \covidscore\ and less noise in whichever users get tested and, as a result of a positive test, get quarantined. As such, the optimization objective is to minimize the fraction $\frac{a}{\rho}$. The constraint for this problem is that $\varepsilon$ is fixed non-linearly for a given $\delta$, $\rho$, and $a$. From~\cite{DBLP:conf/csfw/Mironov17} we know that $\varepsilon = \rho + \frac{\log \frac{1}{\delta}}{a-1}$.

We can write the optimization problem as follows: \\ 

\fbox{\begin{minipage}{15em}
\begin{optim} \label{opt:a}

\[\min_{a,\rho} \frac{a}{\rho}  \]

Such that:
\begin{align}
    \rho + \frac{\log \frac{1}{\delta}}{a-1} - \varepsilon &= 0  \nonumber 
\end{align}
\end{optim}
\vspace{1mm}
\end{minipage}}
\vspace{2mm}

The search space is constrained to $a>1$ and $\rho>0$. For the Lagrangian, $\mathcal{L} = \frac{a}{\rho} + \nu (\rho + \frac{\log \frac{1}{\delta}}{a-1} - \varepsilon) $, we write the optimality conditions~\cite{DBLP:books/cu/BV2014}:
\begin{alignat}{3}
\nabla_a \mathcal{L} =     \frac{1}{\rho} + \nu ( \frac{-\log \frac{1}{\delta}}{(a-1)^2})  &= 0 && \qquad \text{stationarity in $a$}\\ 
\nabla_\rho \mathcal{L} =  \frac{-a}{\rho^2} + \nu  &= 0 && \qquad \text{stationarity in $\rho$} \\ 
\rho + \frac{\log \frac{1}{\delta}}{a-1} &= \varepsilon && \qquad \text{Primal feasibility}   \label{eqn:primal_constraint}
\end{alignat}

For clarity, we rewrite the constant $d = \log \frac{1}{\delta}$. We know $d > 0$, which will be used later. The system of equations results:
\begin{align}
 \frac{1}{\rho} &= \nu ( \frac{d}{(a-1)^2}) \label{eqn:stat_a} \\ 
\frac{a}{\rho^2} &= \nu   \label{eqn:stat_a_rho}\\ 
\rho  &= \varepsilon- \frac{d}{a-1}. \label{eqn:primal_feas} 
\end{align}

Substituting Equation~\ref{eqn:stat_a} in Equation~\ref{eqn:stat_a_rho} yields: 

\begin{align}
    a \frac{\nu^2 d^2}{(a-1)^4} &= \nu.
\end{align}

This means that either $\nu=0$, which is infeasible~\cite{DBLP:books/cu/BV2014}, or 
\begin{align}
   \nu &= \frac{ (a-1)^4}{a d^2} \label{eqn:primal_fourth_order}.
\end{align}

Substitute Equation~\ref{eqn:primal_fourth_order} in Equation~\ref{eqn:stat_a_rho} with Equation~\ref{eqn:primal_feas}:

\begin{align}
    a &= \nu \rho^2 = \frac{ (a-1)^4}{a d^2} (\varepsilon - \frac{d}{a-1})^2 \\ 
    a &=  \frac{ (a-1)^2}{ad^2} (\varepsilon (a-1) - d)^2 \label{eqn:opt_a_use_a_larger_1} \\ 
    a^2 d^2 &=  (a-1)^2 (\varepsilon (a-1) - d)^2  \label{eqn:opt_a_polynomial} 
\end{align}

In Equation~\ref{eqn:opt_a_use_a_larger_1} we use that the order $a$ is larger than 1. \\ 

Equation~\ref{eqn:opt_a_polynomial} is a fourth-order polynomial in $a$. For clarity, we rename $x = a-1$:

\begin{align}
    (x^2 + 2x + 1)d^2 &= x^2(\varepsilon x - d)^2 \\ 
    d^2x^2 + 2d^2x + d^2 &= x^2(\varepsilon^2 x^2 - 2 \varepsilon d x + d^2) \\ 
    d^2x^2 + 2d^2x + d^2 &= \varepsilon^2 x^4 - 2 \varepsilon d x^3 + d^2x^2 \\ 
    \varepsilon^2 x^4 -2 d \varepsilon x^3 & -2d^2x = d^2
\end{align}

This polynomial could be rewritten.
\begin{align}
    (d + \varepsilon x^2) (\varepsilon x^2 - 2 d x - d) = 0
\end{align}

This polynomial has zeroes:~
\begin{alignat}{3}
    x_1 &= i \frac{\sqrt{d}}{\sqrt{\varepsilon}}  \\ 
    x_2 &= -i \frac{\sqrt{d}}{\sqrt{\varepsilon}} \\ 
    x_3 &= \frac{d + \sqrt{d (d+\varepsilon)}}{\varepsilon}  \\ 
    x_4 &= \frac{d - \sqrt{d (d+\varepsilon)}}{\varepsilon}   
\end{alignat}

Variable $i$ indicates the imaginary number. Solutions $x_1$ and $x_2$ are imaginary and have the real part being 0, which is primal infeasible.  Solution $x_4$ is always negative for $\varepsilon>0$, $\delta>0$, which is primal infeasible. Therefore, we use solution $x_3$ and arrive at:

\begin{align}
    a = 1 + \frac{d + \sqrt{d (d+\varepsilon)}}{\varepsilon}
\end{align}

The value for $\rho$ follows from Equation~\ref{eqn:primal_constraint}.

\begin{figure}[t]
    \centering
    \includegraphics[width=0.8\linewidth]{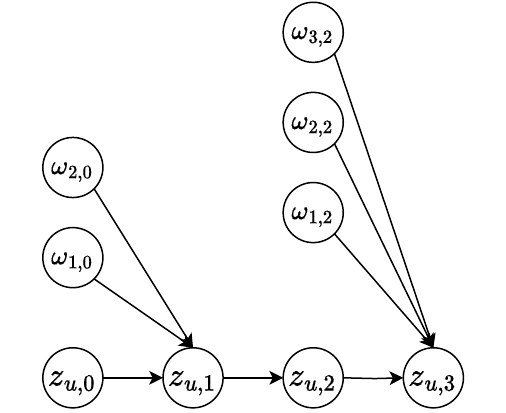}
    \caption{A contact graph with five contacts on two days, as discussed in Appendix Section~\ref{app:fn_traces_explicit}.}
    \label{fig:only_sir}
\end{figure}

\subsection{FN as a Function of Products}\label{app:fn_traces_explicit}

This subsection writes the FN update equations to show that FN can be considered a function with inputs of only the product of messages. First, we work out the updates for a particular small graph. Second,  we generalize this for arbitrary contact graphs. \cite{romijnders2022notimetowaste} derived the update equation for an SEIR model and a general number of timesteps and contacts. For illustration, we write the update equations for a SIR model on four timesteps, with two contacts at day 0, $\omega_{1,0}$ and $\omega_{2,0}$, and with three contacts on day 2, $\omega_{1,2}$ and $\omega_{2,2}$ and $\omega_{3,2}$. The contact graph is illustrated in Figure~\ref{fig:only_sir}. As \covidscore\ for this example, we consider the belief of being in state $I$ at day $t=3$ which is indicated by $b(z_3 = I)$. The user $u$ subscripts are omitted for clarity, i.e., $z_t$ refers in the following to $z_{u,t}$.

Equation~\ref{eqn:explicit01} is the general FN update from~\cite{DBLP:conf/uai/Rosen-ZviJY05} and Equation~\ref{eqn:explicit02} writes out the conditional distribution using the product rule for probabilities. Equation~\ref{eqn:explicit03} simplifies the summation over random variables. As the Markov chain only considers transitions $S \rightarrow I \rightarrow R$, we can reduce computation from the sum that grows exponentially to a sum over traces that grows quadratically or cubically for an SEIR model. This observation has been made in~\cite{crisp}. Equation~\ref{eqn:explicit_pop_sum} writes the summation in full.

\small
\begin{align}
    & b(z_3 = I ) \nonumber \\ 
    &=  \mathbb{E}_{B_{N(u)}} [p(z_3 = I | z_{N(u)})] \label{eqn:explicit01} \\ 
    &=  \mathbb{E}_{B_{N(u)}} [\sum_{z_2} \sum_{z_1} \sum_{z_0} p(z_3 = I| z_2, z_{N(u)}) \nonumber \\ 
    & \quad \cdot p(z_2| z_1, z_{N(u)})p(z_1|z_0, z_{N(u)})p(z_0)] \label{eqn:explicit02} \\ 
    &= \sum_{(z_0,z_1,z2) \in \{ SSS, SSI, SII, III \}} \mathbb{E}_{B_{N(u)}} [ p(z_3 = I| z_2, z_{N(u)})\nonumber \\ 
    & \quad \cdot p(z_2|z_1, z_{N(u)})p(z_1|z_0)p(z_0)] \label{eqn:explicit03} \\ 
    & = \mathbb{E}_{B_{N(u)}} [p(z_3 = I| z_2=S, z_{N(u)}) \nonumber \\ 
    & \quad \cdot p(z_2 = S| z_1=S) p(z_1=S|z_0=S, z_{N(u)}) p(z_0=S)] \nonumber \\
    & \qquad +  \mathbb{E}_{B_{N(u)}} [ p(z_3 = I| z_2=I, z_{N(u)}) \nonumber \\ 
    & \quad \cdot p(z_2 = I| z_1=S) p(z_1=S|z_0=S, z_{N(u)}) p(z_0=S)]  \nonumber \\
    & \qquad +  \mathbb{E}_{B_{N(u)}} [ p(z_3 = I| z_2=I, z_{N(u)}) \nonumber \\ 
    & \quad \cdot p(z_2 = I| z_1=I) p(z_1=I|z_0=S, z_{N(u)}) p(z_0=S)]  \nonumber \\
    & \qquad +  \mathbb{E}_{B_{N(u)}} [ p(z_3 = I| z_2=I, z_{N(u)}) \nonumber \\ 
    & \quad \cdot p(z_2 = I| z_1=I) p(z_1=I|z_0=I, z_{N(u)}) p(z_0=I)] \label{eqn:explicit_pop_sum} \\ 
    & = \mathbb{E}_{B_{N(u)}} [p(z_3 = I| z_2=S, z_{N(u)}) \nonumber \\ 
    & \quad \cdot (1-p_0) p(z_1=S|z_0=S, z_{N(u)})  (1-p_0)] \nonumber \\
    & \qquad +  \mathbb{E}_{B_{N(u)}} [ g p_0 p(z_1=S|z_0=S, z_{N(u)}) (1-p_0)]  \nonumber \\
    & \qquad +  \mathbb{E}_{B_{N(u)}} [ g g p(z_1=I|z_0=S, z_{N(u)}) (1-p_0)]  \nonumber \\
    & \qquad +  \mathbb{E}_{B_{N(u)}} [ g g g p_0] \label{eqn:explicit_known_scalars} \\ 
    & = \left(1-(1-p_0)\textcolor{blue}{(1-p_1\phi_{1,2})(1-p_1\phi_{2,2})(1-p_1\phi_{3,2})}\right) \nonumber \\ 
    & \quad \cdot (1-p_0) (1-p_0)\textcolor{red}{(1-p_1\phi_{1,0})(1-p_1\phi_{2,0})}  (1-p_0) \nonumber \\
    & \qquad +   g p_0 (1-p_0)\textcolor{red}{(1-p_1\phi_{1,0})(1-p_1\phi_{2,0})} (1-p_0)  \nonumber \\
    & \qquad +   g g (1-(1-p_0)\textcolor{red}{(1-p_1\phi_{1,0})(1-p_1\phi_{2,0})}) (1-p_0)  \nonumber \\
    & \qquad +   g g g p_0 \label{eqn:dpfn_four_lines} \\
    & = (1-(1-p_0)\textcolor{blue}{\omega_{1,2}\omega_{2,2}\omega_{3,2})}) \nonumber \\ 
    & \quad \cdot (1-p_0) (1-p_0)\textcolor{red}{\omega_{1,0}\omega_{2,0}}  (1-p_0) \nonumber \\
    & \qquad +   g p_0 (1-p_0)\textcolor{red}{\omega_{1,0}\omega_{2,0}} (1-p_0)  \nonumber \\
    & \qquad +   g g (1-(1-p_0)\textcolor{red}{\omega_{1,0}\omega_{2,0}}) (1-p_0)  \nonumber \\
    & \qquad +   g g g p_0. \label{eqn:dpfn_four_lines_S} 
\end{align}
\normalsize

In Equation~\ref{eqn:explicit_known_scalars} we replace the known scalars for prior belief $p_0$, and probability $p(z_{t+1} = I| z_t=I, z_{N(u,t)}) = g$. In Equation~\ref{eqn:dpfn_four_lines}, we replace the conditional distribution under the local belief according to Equation 8 of~\cite{romijnders2022notimetowaste}.

Equation~\ref{eqn:dpfn_four_lines_S} makes explicit that the output from FN only depends on a product of scores. From Equation~\ref{eqn:dpfn_four_lines} we write the messages in shorthand. If the user receives a message $\phi_{c, t}$ from contact $c$ on time $t$, then rewrite the message to: 

\begin{equation} \label{eqn:phi_to_S}
  \omega_{c, t} = 1 - p_1 \phi_{c,t}.  
\end{equation}
~With this transform, FN depends on products of the $\omega_{c,t}$.

While this example was specific for an SIR in the graph in Figure~\ref{fig:only_sir}, previous work showed the FN update for an SEIR model in general. In full generality, the messages, $\omega_{c,t}$, occur in each conditional probability distribution and can in each case be rewritten to the product $\omega_{*,t}$, similar to Equation~\ref{eqn:dpfn_four_lines_S}. Such product term would only ever appear in a conditional distribution $p(z_{u,t+1}| z_{u,t}=S, z_{N(u)})$, which are the transitions $S \rightarrow S$ or $S \rightarrow E$. Observations appear as a distinct factor and do not influence the above product. Therefore, FN on a contact graph with arbitrary contacts and arbitrary observations can be rewritten to depend on the product $\omega_{*,t}$, and Algorithm~\ref{alg:dpfn} describes its differentially private implementation.

\subsection{Experimental Details} \label{app:experimental_details}

In addition to the experimental details in Section~\ref{sec:experimental_details}, we expand on a few more points pertaining to our experiments. 

The most important difference between the \covasim\ simulator and the experiments using \openabm\ is that in \covasim, each day 10\% of the population may be tested and in \openabm, each day 2\% of the population may be tested. We found that in \covasim, when testing too many agents daily, all methods would achieve an equally low peak infection rate, i.e. no pandemic. Such setting would not allow for a comparison of the methods. In both simulators, isolation is for ten days, and the first isolation only starts on the fourth day; the first three days are used to start the simulation.

The simulator settings for \openabm\ follow previous literature~\cite{baker2021epidemic,romijnders2022notimetowaste}. Each simulation starts with 25 users in infectious state; scale experiments with 500.000 and more users start with 50 users in infectious state. The simulator settings for \covasim\ follow the recommended tutorials. The simulator is set with dates February 1st 2020 to May 1st 2020, which accounts for 91 days (whereas \openabm\ runs for 100 days). Following recommendations by the authors, population type `hybrid' is used.  Each simulation starts with 25 users in infectious state; scale experiments with 500.000 and more users start with 100 users in infectious state.

The \openabm\ simulator uses ten health states, Uninfected (0), Pre-symptomatic (1), Mild pre-symptomatic (2), Asymptomatic (3), Symptomatic (4), Mild symptomatic (5), Hospitalised (6), Critical (7), Recovered (8), Death (9). A user can test positive from state one to seven. We report infection rate as occurrence of state three to seven. The \covasim\ simulator models with five states, Susceptible, Exposed, Infectious (which is a subset of Exposed), Recovered, and Death. A user can test positive in the Exposed state. We report infection rate as the occurrence of the Infectious state.

The model, as defined in Section~\ref{sec:method} has four model parameters, and we have set them based on insights from the literature.  Model parameters are set at $p_0=\frac{1}{1000}$, $p_1=0.05$, $g=0.99$, $h=0.10$. This can be interpreted as that the typical infection lasts eleven days, where the user is not considered to spread the virus on the first day of infection. Only for the visualization in Figure~\ref{fig:fn_analysis_score}, the parameter $p_1$ is exaggerated to $0.25$ for illustration purposes.

Parameters specific to each method. In Gibbs sampling, we take ten samples, each separated by ten skip steps, and each chain is burned in for 100 steps. The likelihoods are clipped at a norm of 10. In per-message differential privacy, the messages are clipped between $[0.01, 0.99]$, and the Gaussian mechanism is run in the logit domain\footnote{An earlier version of this paper incorrectly used the Gaussian Mechanism for all $\varepsilon$. In this version, Figure~\ref{fig:pareto} has been rerun using the Analytical Gaussian Mechanism from~\citet{balle2018gauss} for $\varepsilon>1$ and the results stay comparable to before.}. 
In the \textit{Traditional contact tracing }method, a sampled \covidscore\ could be negative due to the tails of the Gaussian distribution. However, we clip negative values to zero as it is public knowledge that a \covidscore\ must be non-negative. Experiments with FN use the aforementioned model parameters, and assumes that the false positive and false negative rates are known. In future, one could experiment with different noise rates between noisier tests, e.g. self testing, and less noisy tests, e.g. testing in lab environments. 

All experiments run inference in a time window of fourteen days. This setting is in accordance with previous works~\cite{baker2021epidemic,romijnders2022notimetowaste}. The prior for each window is similar to~\cite{crisp}: $p(z_{u,0} = S) = 1-p_0$, $p(z_{u,0} = E) = p_0$, and $p(z_{u,0}=I) = p(z_{u,0}=R) = 0$. 

\begin{table}[t]
    \centering
    \setlength{\tabcolsep}{4pt}
    \begin{tabularx}{\linewidth}{l|l}
      \textbf{Hyperparameter setting} & \textbf{PIR} (\permil) \Bstrut \\
      \hline  
    number of days = \num{7}    & $133.5_{ \ \ [131.8,137.3]}$ \Tstrut \\
    number of days = \num{10}   & $17.9_{ \ \ [12.2,21.1]}$  \\
    number of days = \num{14}   & $1.0_{ \ \ [0.8,1.3]}$  \\
    number of days = \num{21}   & $1.0_{ \ \ [0.9,1.5]}$  \\
    number of days = \num{30}   & $1.1_{ \ \ [0.9,1.4]}$ \Bstrut \\
    \hline    
 $p_1 = \ $\num{0.001} &  $3.9_{ \ \ [1.7,7.6]}$ \Tstrut \\ 
 $p_1 = \ $\num{0.010} &  $1.1_{ \ \ [0.9,1.7]}$ \\ 
 $p_1 = \ $\num{0.050} &  $1.0_{ \ \ [0.8,1.3]}$ \\ 
 $p_1 = \ $\num{0.100} &  $1.0_{ \ \ [0.9,1.2]}$ \\ 
    \end{tabularx}
    \caption{Two hyperparameter sweeps for important parameters of our algorithm. Numbers indicate the median peak infection rate (PIR) as one infection per thousand (\permil). The subscripts indicate 20-80 quantiles of ten random restarts. }
    \label{tab:hyper_numdays}
\end{table}

\textbf{Decentralization: } All methods that we study in this paper are decentralized. No (central) entity knows about the \covidscore\ of multiple individuals. Both DPFN, Gibbs sampling, and Traditional contact tracing operate by sending messages between users. One step in our simulation that needs further clarification is the selection of which people to signal for a \covidn\ test. Currently, the simulation selects the users with the highest \covidscore\ for a test. This computation can happen encrypted~\cite{beaver1990round,ben2016optimizing}, or differentially private~\cite{dwork_dp}. A different scenario that one could experiment with is that the tests are based on a threshold of the \covidscore\, but making more tests reduces PIR, so this makes the results hard to compare. Therefore, to compare the contact tracing algorithms, we use the ranking approach for the testing policy.

\textbf{Hyperparameters: } During this research, we have made hyperparameter sweeps for two important parameters, and we report the results of that sweep here. 

The duration of the inference window changes how many days of past contacts influence a particular \covidscore. Increasing this window makes the statistical inference more accurate but would create a possibly worse composition for privacy and increased computing costs. The computation scales cubically with the size of the window~\cite{romijnders2022notimetowaste}. Table~\ref{tab:hyper_numdays} displays the result of this sweep, which is run with $\varepsilon=1$. We have chosen the value of fourteen, which strikes a balance between the privacy and compute cost on the one hand and the PIR on the other hand.

The second hyperparameter sweep concerns the value of $p_1$ in our experiments. This parameter is paramount, as it appears in both the statistical model in Equation~\ref{eqn:crisp_noisy_or} and the privacy bound in Equation~\ref{eqn:bound_variance_rdp}. The value is estimated by population studies at 0.01~\cite{openabm}, and misspecifying the value has the problem of model misspecification during inference of the \covidscore. On one side, a low value of $p_1$ requires less noise to satisfy DP per Equation~\ref{eqn:bound_variance_rdp}. On the other side, a slightly higher value achieves a lower, better peak infection rate. We hypothesize this effect is due to a higher value of $p_1$ emphasizing contact tracing and introducing a stronger connection between a user and its contacts' scores. Table~\ref{tab:hyper_numdays} displays the results of this sweep at $\varepsilon=1$ and window size 14. We pick a value $p_1=0.05$, which has the lowest median PIR, also taking into consideration the quantiles, which are slightly lower for $p_1=0.05$. 

\textbf{Technical details: } 
For the \openabm\ simulator, we use code at \url{github.com/BDI-pathogens/OpenABM-Covid19}. For the \covasim\ simulator, we use code at \url{github.com/InstituteforDiseaseModeling-/covasim}. The parameter files for \openabm\ can be found at \url{github.com/BDI-pathogens/OpenABM-Covid19/tree/master/tests/data/*.csv}. 

Our experiments run on a 32-core CPU node with less than 60GB of memory. Only the experiments with over a half million users require a compute node with 120GB memory or more. Experiments with 100.000 users generally finish in under three hours each. We are also releasing an implementation in C++, which finishes such experiment in about 30 minutes for 16 parallel simulations. The experiments with belief propagation generally take ten times longer, which amounts to about a whole day for 100.000 users. Due to compound computing requirements in the inference algorithm, the experiments with one million users can take up to forty hours. The python implementation uses Numba runtime jit-compilation~\cite{lam2015numba}.

In Figure~\ref{fig:loss_follow_up}, the experiment for the loss to follow-up directly modulates this parameter in the \covasim\ simulator. This parameter is defined as \texttt{loss\_prob} in \texttt{covasim/interventions.py}.

Python implementation:

\texttt{github.com/RobRomijnders/dpfn\_aaai}.

C++ implementation: 

\texttt{github.com/RobRomijnders/dpfn\_cpp}.
\fi
\end{document}